\documentclass[11pt,3p,times]{elsarticle}

\usepackage{amsmath,amsfonts,amssymb}
\usepackage{amsthm}
\usepackage{bm}
\usepackage{graphicx,epstopdf}
\usepackage[position=top,labelfont=normalfont,textfont=normalfont,singlelinecheck=off,justification=raggedright]{subfig}
\usepackage{floatrow}
\usepackage[dvipsnames]{xcolor}
\usepackage{setspace}
\usepackage{enumerate,etaremune}
\usepackage{booktabs}
\usepackage{tabularx,multirow}
\usepackage{framed}
\usepackage{hhline}
\usepackage{url}
\usepackage{import}
\usepackage[unicode,bookmarks=false]{hyperref}
\hypersetup{
  colorlinks,
  citecolor=Blue,linkcolor=Blue,urlcolor=Blue
}

\usepackage[textsize=footnotesize,linecolor=red,backgroundcolor=red!25,bordercolor=red]
  {todonotes}

\usepackage{tikz}
\usetikzlibrary{shapes.geometric}
\usetikzlibrary{shapes,arrows}
\usetikzlibrary{plotmarks}

\usetikzlibrary{calc}

\usepackage{physics}

\usepackage{algorithm}
\usepackage{algorithmicx,algpseudocode}
\usepackage{cases}
\newcommand{\eg}{{\it e.g.,}}
\newcommand{\ie}{{\it i.e.,}}

\newcommand{\etal}{{\it et al.}}
\newcommand{\tensor}[1]{\bm{#1}}
\newcommand{\stress}{\sigma}
\newcommand{\strain}{\varepsilon}
\newcommand{\tstress}{\tensor{\stress}}
\newcommand{\tstrain}{\tensor{\strain}}

\newcommand{\pd}{\partial}

\newcommand{\rn}[1]{\uppercase\expandafter{\romannumeral #1\relax}}

\let\diver\relax
\DeclareMathOperator{\diver}{\nabla\cdot}

\DeclareMathOperator{\vol}{vol}

\newsavebox{\dotbox}

\theoremstyle{remark}

\newcolumntype{L}[1]{>{\raggedright\let\newline\\arraybackslash\hspace{0pt}}m{#1}}
\newcolumntype{C}[1]{>{\centering\let\newline\\arraybackslash\hspace{0pt}}m{#1}}
\newcolumntype{R}[1]{>{\raggedleft\let\newline\\arraybackslash\hspace{0pt}}m{#1}}

\allowdisplaybreaks

\biboptions{sort&compress,square,comma,numbers}
\AtBeginDocument{\hypersetup{citecolor=MidnightBlue,linkcolor=MidnightBlue,urlcolor=MidnightBlue}}

\usepackage{etoolbox}
\makeatletter
\patchcmd{\ps@pprintTitle}% <cmd>
{Preprint submitted to}% <search>
{~}% <replace>
{}{}% <succes><failure>
\makeatother
\begin{document}

\begin{frontmatter}

\title{A phase-field model for hydraulic fracture nucleation and propagation in porous media}

\author[LLNL]{Fan Fei\corref{corr}}
\cortext[corr]{Corresponding Author}
\ead{fei2@llnl.gov}
\author[DUKE]{Andre Costa}
\author[DUKE]{John E. Dolbow}
\author[LLNL]{Randolph R. Settgast}
\author[LLNL]{Matteo Cusini}

\address[LLNL]{Atmospheric, Earth, and Energy Division, Lawrence Livermore National Laboratory, USA}
\address[DUKE]{Department of Mechanical Engineering and Materials Science, Duke University, Durham, USA}

\journal{~}

\begin{abstract}

 Many geo-engineering applications, \eg~enhanced geothermal systems, rely on hydraulic fracturing to enhance the permeability of natural formations and allow for sufficient fluid circulation.  
Over the past few decades, the phase-field method has grown in popularity as a valid approach to modeling hydraulic fracturing because of the ease of handling complex fracture propagation geometries. However, existing phase-field methods cannot appropriately capture nucleation of hydraulic fractures because their formulations are solely energy-based and do not explicitly take into account the strength of the material. Thus, in this work, we propose a novel phase-field formulation for hydraulic fracturing with the main goal of modeling  fracture nucleation in porous media, \eg~rocks. 
Built on the variational formulation of previous phase-field methods, the proposed model incorporates the material strength envelope for hydraulic fracture nucleation through two important steps: (i) an external driving force term, included in the damage evolution equation, that accounts for the material strength; (ii) a properly designed damage function that defines the fluid pressure contribution on the crack driving force. 
The comparison of numerical results for two-dimensional (2D) test cases with existing analytical solutions demonstrates that the proposed phase-field model can accurately model both nucleation and propagation of hydraulic fractures. Additionally, we present the simulation of hydraulic fracturing in a three-dimensional (3D) domain with various stress conditions to demonstrate the applicability of the method to realistic scenarios.   

\end{abstract}

 
\end{frontmatter}

% \linenumbers

% SECTION 1
% ------------------------------------------------------------------------------
\section{Introduction}
\label{sec:intro}

%----------------------------------------------------------------------------------------
%	INTRODUCTION
%----------------------------------------------------------------------------------------

Hydraulic fracturing consists in enhancing the permeability of a natural formation by injecting a fracturing fluid (\eg~water) at a high pressure. This technique has been widely used in many geo-engineering applications, such as unconventional oil and gas production~\cite{economides1989reservoir,warpinski2009stimulating,cueto2013forecasting} and enhanced geothermal systems (EGS)~\cite{brown2012mining,legarth2005hydraulic,mcclure2014investigation}, in which rock masses are nearly impermeable. For EGS, the effectiveness of the stimulation treatment determines the performance of a target site. Thus, being able to thoroughly understand the hydraulic fracturing process is crucial to operate such systems safely and effectively. As a consequence, there has been a growing interest in the development of numerical approaches to model hydraulic fracturing. 

A popular modeling choice is to consider fractures as sharp interfaces, represented by lower dimensional manifolds embedded in a higher dimensional domain (\eg~surfaces in a 3D domain). To this end, there exist several approaches to include discontinuities in both finite element (FE) and finite volume (FV) methods for flow, mechanics, and poromechanics models. These approaches can be grouped into two main classes: (i) conforming methods, in which fractures are represented by 2D elements that coincide with the boundaries of 3D cells, \eg~\cite{fu2013explicitly,settgast2017fully}; (ii) embedded approaches, in which fractures are meshed independently of the rock matrix domain and the formulation is enriched to account for their effects, \eg~\cite{moes1999finite,lecampion2009extended,mohammadnejad2013extended,gupta2014simulation}. Both these approaches suffer from some limitations and involve several practical challenges when employed to model hydraulic fracturing. For example, conforming approaches only allow fractures to propagate along element boundaries (faces in 3D), which prevents fractures from growing in an arbitrary direction. While embedded approaches can be used to overcome such limitation, they often face challenges in defining general methods to identify the direction of fracture propagation and handling complex geometries (\eg~crack branching). 

Over the past few decades, the phase-field model for fractures has been identified as a promising alternative to sharp interface approaches to model fracture propagation in rocks and rock-like materials~\cite{zhang2017modification,choo2018coupled,fei2020phase,fei2021double}. The fracture, instead of being explicitly represented as an interface, is approximated by a diffuse variable (\ie~damage). The main advantage of this diffuse crack representation is the simplicity of representing complex geometries. There have also been several efforts to extend the phase-field method to model hydraulic fracturing~\cite{mikelic2015phase,wilson2016phase,lee2016pressure,santillan2018phase,chukwudozie2019variational} in poroelastic media. Hybrid methods that combine sharp and diffuse crack representations to leverage the strengths of both approaches have also been proposed recently~\cite{costa2022,zhang2022hybrid}. 

Most of the methods developed so far, based on either sharp or diffuse crack representations, have focused on the propagation of preexisting hydraulic fractures, while little attention has been given to modeling fracture nucleation in bulk materials. However, understanding hydraulic fracture nucleation in the near-wellbore region can provide crucial information about the in-situ stress~\cite{zoback2010reservoir} and drive the design of more efficient hydraulic fracturing operations~\cite{huang2012initiation}. 
Existing phase-field methods for hydraulic fracturing are mostly based on a variational formulation that casts the fracture propagation problem in terms of the minimization of total potential energy of the system~\cite{wheeler2014augmented,mikelic2015quasi,mikelic2015phase,santillan2018phase,chukwudozie2019variational}. This formulation can accurately model fracture propagation, in good agreement with the classic fracture mechanics theory as originally described in~\cite{griffith1921vi}. However, casting the problem solely in terms of energy minimization completely neglects the material strength and how fractures may nucleate in the bulk due to a stress-induced failure. Additionally, this purely energetic formulation results in a dependency of the material strength on the phase-field regularization length, which is a parameter that governs the width of the diffuse fracture region. This length dependency issue has been thoroughly discussed in the literature (see~\eg~\cite{wu2017unified,geelen2019phase,mandal2019length,fei2020phase}). 
As a consequence, the regularization length becomes a material-specific parameter that should be calibrated based on the tensile and compressive strengths of a given material~\cite{zhang2017numerical,choo2018coupled}. 
%The calibrated regularization length can be extremely small compared to the scale of interest. Consequently, one must use very high resolution meshes which are often impractical for subsurface applications even with today's computational resources.
This length dependency issue has been addressed in the recent works on phase-field models for quasi-brittle materials~\cite{wu2017unified,geelen2019phase,fei2020phase}, which are inspired by a gradient damage model introduced by Lorentz and Godard~\cite{lorentz2011gradient}.
Unfortunately, these models are still derived through an energy minimization process, and as such damage nucleation is governed by a threshold that is energetic.  
In a recent work, Kumar~\etal~\cite{kumar2020revisiting} have proposed a novel phase-field formulation to address these issues. In essence, they have included a well-designed external driving force term in the phase-field formulation to account for the assumption of a strength envelope governing nucleation. When correctly designed, this additional term allows the resulting phase-field model to approximate the strength surface of the material. To date, the work of Kumar~\etal~\cite{kumar2020revisiting} has been restricted to traction-free crack surfaces.  An analogous model that accounts for pressurized cracks and nucleation in porous media has yet to be developed.  

In this work, we present a novel phase-field formulation to model nucleation and propagation of hydraulic fractures in porous materials. Specifically, we extend the formulation of the previous phase-field model for hydraulic fracturing by including the external driving force term proposed by Kumar~\etal~\cite{kumar2020revisiting}. To ensure that the model correctly reproduces the material strength even in the presence of fluid pressure, we devise a special damage function for the pressure terms in the phase-field formulation. The proposed model accurately predicts hydraulic fracture nucleation in an intact porous material while retaining the ability to model fracture propagation.

% Review at the end
The paper is organized as follows. In Section~\ref{sec:formulation}, we review the variational phase-field method and its formulation for hydraulic fracturing. In Section~\ref{sec:nucleation} we extend the phase-field formulation presented in Section~\ref{sec:formulation} by adding an external driving force and devising the proper damage function for pressure-dependent terms. The discretization and the solution strategy are described in Section~\ref{sec:numerical} whereas, in Section~\ref{sec:simulation}, we present a series of two-dimensional (2D) and three-dimensional (3D) numerical examples to demonstrate the model correctly represents hydraulic fracture nucleation and propagation. The paper is concluded in Section~\ref{sec:conclusion}.

% SECTION 2
% ------------------------------------------------------------------------------
\section{The phase-field model for hydraulic fracture propagation}
\label{sec:formulation}

%----------------------------------------------------------------------------------------
%   FORMULATION
%----------------------------------------------------------------------------------------

The goal of this section is to review the phase-field formulation for hydraulic fracturing in saturated porous materials (\eg~rocks). 

\subsection{Problem statement}
\begin{figure}[htbp]
    \centering
    \includegraphics[height=0.3\textwidth]{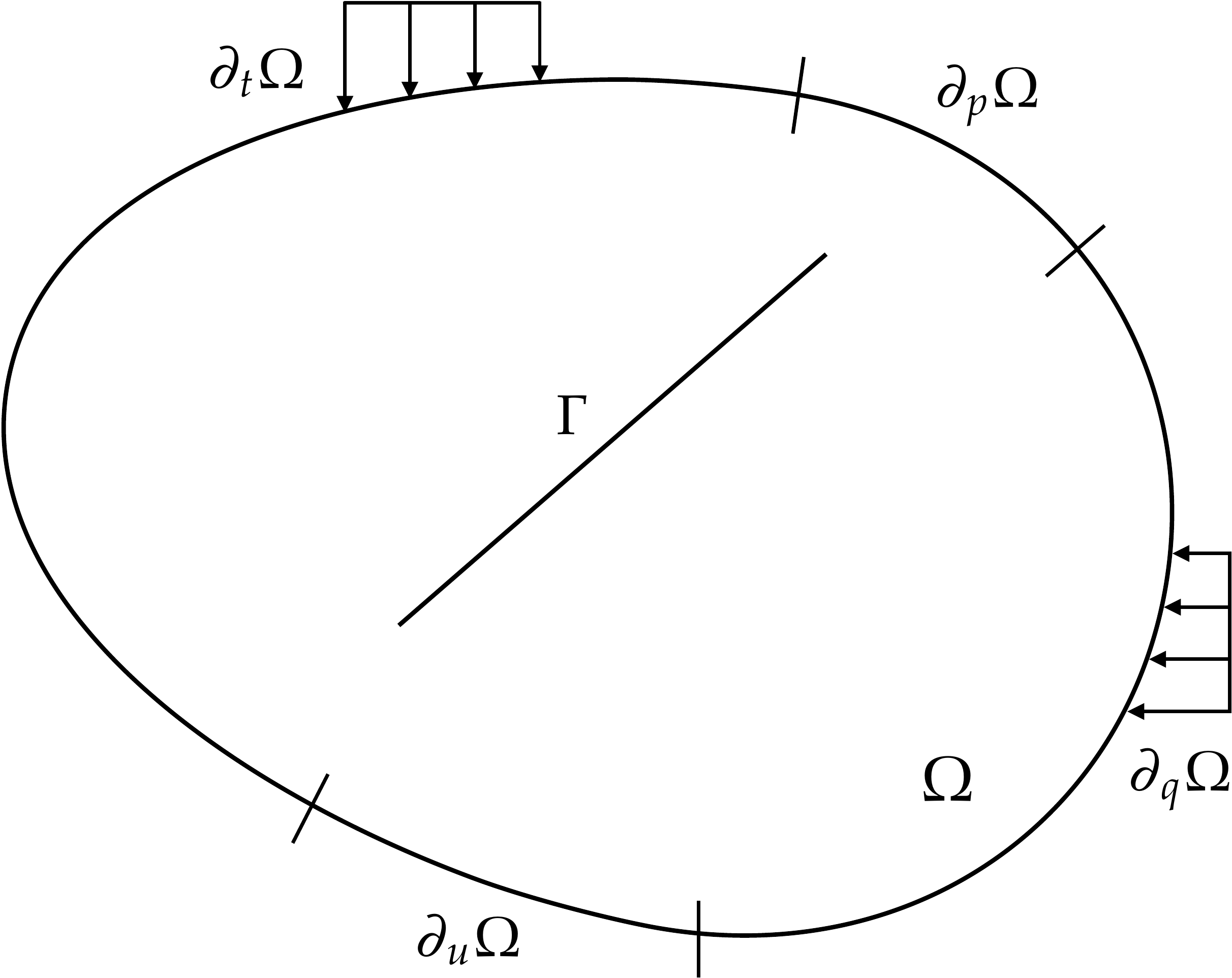}
    \caption{Problem geometry.}
    \label{fig:problem-state}
\end{figure}

Let us consider a porous medium $\Omega \in \mathbb{R}^{n_\mathrm{dim}}$ in Figure~\ref{fig:problem-state}, with $n_\mathrm{dim}$ denoting the spatial dimension. The external boundary, $\pd \Omega$, is divided into non-overlapping portions $\pd_{u} \Omega \cup \pd_{t} \Omega = \pd_{p} \Omega \cup \pd_{q} \Omega = \pd \Omega $, identifying where Dirichlet and Neumann boundary conditions for mechanics and flow problems will be applied. The continuous domain encloses a set of fractures identified by the lower dimensional domain $\Gamma$. For simplicity, we assume that fractures do not intersect the external boundary. Both the porous medium and fractures are fully saturated by a single-phase Newtonian fluid. Given the initial state, the goal is to model the system evolution in terms of fluid pressure ($p$), mechanical deformation ($\tensor{u}$), and fracture geometry ($\Gamma$) over the time domain $\mathbb{T} = (0, t_\text{max}]$. Based on the variational approach for fractures in poroelastic media~\cite{mikelic2015quasi,santillan2018phase}, the system evolves such that the total potential energy of the system is minimized. The total potential energy is a function of the deformation field, the fluid pressure in both rock matrix and fractures, and the fracture geometry, \ie
\begin{align}
    \label{eq:tpe}
    \Psi(\tensor{u}, p, \Gamma) = \Psi^{e}(\tstrain, \Gamma)  + \Psi^{p}(\tensor{u}, p, \Gamma) + \Psi^{d}(\Gamma) - \Psi^{s}(\tensor{u}, \Gamma).
\end{align}
Here, $\Psi^{e}$ is the elastic strain energy of the bulk solid, $\Psi^{p}$ is the energy stored in the fluid, $\Psi^{d}$ is the fracture dissipation, and $\Psi^{s}$ is the work of external forces. The minimization of the total potential energy is coupled with the mass balance equation describing single-phase flow in a porous medium. 

\subsection{The phase-field regularization}
\begin{figure}[htbp]
    \centering
    \includegraphics[height=0.3\textwidth]{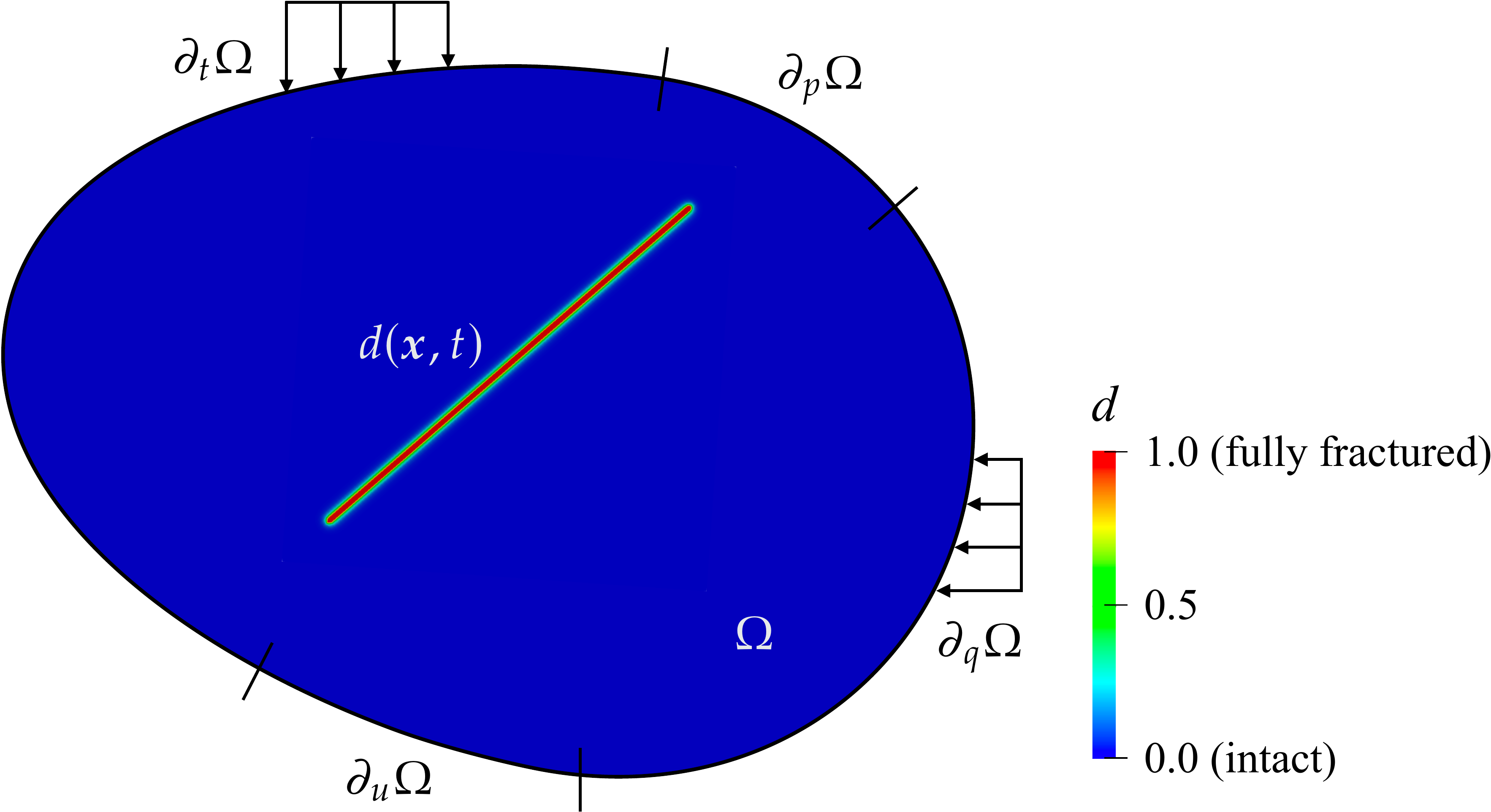}
    \caption{Phase-field approximation of hydraulic fractures.}
    \label{fig:phase-field-approx}
\end{figure}

The phase-field method approximates the fracture geometry, $\Gamma$, by introducing a continuous variable, the damage ($d$), bounded between 0 (intact rock) and 1 (fully fractured) as shown in Figure~\ref{fig:phase-field-approx}, and a regularization length, $L$, that defines the width of the diffuse region. As such, the total energy of the system is approximated by
\begin{align}
\Psi(\tensor{u}, p, \Gamma ) \approx \tilde{\Psi}(\tensor{u}, p, d) = \tilde{\Psi}^{e}(\tstrain, d)  + \tilde{\Psi}^{p}(\tensor{u}, p, d) + \tilde{\Psi}^{d}(d) - \tilde{\Psi}^{s}(\tensor{u}, d).
\end{align} 

Given the phase-field approximation, the elastic strain energy $\Psi^{e}$ reads
\begin{align}
    \Psi^{e} (\tstrain, \Gamma)  \approx \tilde{\Psi}^{e}(\tstrain, d)  = \int_{\Omega} g(d) W^{e} (\tstrain) \: \dd V.  \label{eq:psi-elastic}
\end{align}
Here, $W^{e}(\tstrain)$ is the strain energy of the intact material, \ie
\begin{align}
    W^{e}(\tstrain) = \dfrac{1}{2}\tstrain : \mathbb{C}^{e}:\tstrain,
\end{align}
where $\tstrain :=  \grad^{\mathrm{s}} \tensor{u}$ is the strain tensor and $\mathbb{C}^{e}$ is a fourth-order tensor of elastic moduli. $g(d)$ is, instead, a smooth function that has the effect of degrading the material stiffness whenever damage is non-zero. In this work, we employ a quadratic form of $g(d)$, as done in several other works~\cite{miehe2010thermodynamically,miehe2010phase,borden2012phase}, \ie
\begin{align}
    g(d) = (1 - d)^2. 
\end{align}

The potential energy of the fluid $\Psi^{p}$ can be approximated by a volume integral~\cite{mikelic2015quasi,santillan2018phase}, \ie
\begin{align}
   \Psi^{p}(\tensor{u}, p, \Gamma) \approx \tilde{\Psi}^{p}(\tensor{u}, p, d) = \int_{\Omega} - m(d) p b \diver  \tensor{u} \: \dd V,  \label{eq:psi-fluid}
\end{align}
where $b$ is the Biot's coefficient and $m(d)$ is a function that satisfies the following constraints,
\begin{align}
    m(0) = 1; \, m(1) = 0; \, m'(1) = 0; \, m(d) < 0, \,\,\text{when} \,\, 0 < d < 1. \label{eq:md-requirement}
\end{align}

The energy dissipated by the fracture $\Psi^{d}$ is defined as the integral of the critical fracture energy $\mathcal{G}_{c}$ over the fracture surface, and we approximate it as follows in the phase-field model~\cite{wu2017unified,geelen2019phase,fei2020phase},
\begin{align}
   \Psi^{d} =  \int_{\Gamma} \mathcal{G}_{c} \: \dd A \approx \tilde{\Psi}^{d} = \int_{\Omega} \mathcal{G}_{c}\dfrac{1}{c_{0}L} \left[ \omega(d) + L \grad d \cdot \grad d\right] \: \dd V \, \, \text{with} \, \, c_{0} = 4\int_{0}^{1} \sqrt{\omega(s)} \: \dd s , \label{eq:psi-fracture}
\end{align}
where $\omega(d)$ is a local dissipation function. In most existing phase-field models for hydraulic fracturing, the local dissipation function is $\omega(d) = d^2$, \eg~\cite{santillan2018phase,mikelic2015quasi,lee2016pressure,wheeler2014augmented,bourdin2012variational,mikelic2015phase,miehe2016phase}. 
However, this choice leads to damage growth as soon as the strain energy is non-zero.   Therefore, given the goal of modeling strength-based fracture nucleation, we choose a linear form $\omega(d) = d$, as introduced in Pham~\etal~\cite{pham2011gradient} and previously employed in other phase-field studies~\cite{geelen2019phase,fei2020phase}. 

Finally, the total work done by external forces reads
\begin{align}
    \Psi^{s} = \int_{\Omega} \rho \tensor{g} \cdot \tensor{u} \: \dd V + \int_{\pd_{t} \Omega \cup \Gamma} {\tensor{t}} \cdot \tensor{u} \: \dd A ,  \label{eq:psi-ext}
\end{align}
where $\rho$ is the mass density and $\tensor{g}$ is the gravitational vector. Here, the first integral represents the work done by the body force, and the second one is the work of the traction $\tensor{t}$ on the external boundary and on the fracture surface.
The traction on the fracture surface is exerted by the fluid pressure, \ie
\begin{align}
    \int_{\Gamma} {\tensor{t}} \cdot \tensor{u} \: \dd A = - \int_{\Gamma} p \tensor{n} \cdot \tensor{u} \: \dd A, \label{eq:traction-work} 
\end{align}
where $\tensor{n}$ is the unit normal vector pointing outward. By applying the divergence theorem and assuming a homogeneous displacement boundary condition, Eq.~\eqref{eq:traction-work} can be transformed into   
\begin{align}
    \int_{\Gamma} p \tensor{n} \cdot \tensor{u} \: \dd A = \int_{\Omega \backslash \Gamma} \nabla \cdot  (p \tensor{u}) \: \dd V - \int_{\pd_{t} \Omega} p \tensor{n} \cdot \tensor{u} \: \dd A \label{eq:fracture-pressure-work-final}.
\end{align}
Then, introducing the phase-field regularization, the work of external forces is approximated by
\begin{align}
    \Psi^{s}(\tensor{u}, p, \Gamma) \approx \tilde{\Psi}^{s}(\tensor{u}, p, d) = \int_{\Omega} \rho \tensor{g} \cdot \tensor{u} \: \dd V  - \int_{\Omega} m(d) \div(p \tensor{u}) \: \dd V + \int_{\pd_{t} \Omega} p \tensor{n} \cdot \tensor{u} \: \dd A + \int_{\pd_{t} \Omega} \tensor{t} \cdot \tensor{u} \: \dd A.  \label{eq:psi-ext-final}
\end{align}

\subsection{Governing equations}
Given the phase-field regularization introduced in the previous subsection, the momentum balance and the evolution equation for the damage field can be obtained from the minimization of the total potential energy, \ie
\begin{equation}
    \label{eq:minimization}
    \begin{split}
    (\tensor{u}, d ) &= \underset{\tensor{u}, d}{\text{argmin}}\left[\tilde{\Psi}(\tensor{u}, p, d) \right] \quad \text{in} \,\, \Omega \times \mathbb{T} \\
    \text{with} & \quad d \in [0, 1] \,\, \text{and} \, \, \dot{d} \geq 0. 
    \end{split}
\end{equation}
%The momentum balance and damage evolution equations can be derived from Eq.~\eqref{eq:minimization}, based on the Karush–Kuhn–Tucker conditions. 
These equations are further coupled with the mass conservation equation describing single-phase flow in a porous medium. Thus, the strong form of the initial boundary value problem is to find the displacement ($\tensor{u}$), the damage ($d$), and the fluid pressure ($p$) that satisfy
\begin{align}
    \nabla \cdot \left[ \tstress'(\tstrain, d) - m(d) (b-1) p \tensor{1} \right] - m(d) \grad p + \rho \tensor{g}  = \tensor{0} \quad \text{in} & \,\, \Omega \times \mathbb{T}, \\
    \left \{
    \begin{array}{ll}
         2\,(d-1) W^{e}(\tstrain) + m'(d) \left[(1 - b)p \nabla \cdot \tensor{u} + \grad p \cdot \tensor{u} \right] + \dfrac{3\mathcal{G}_{c}}{8 L } \left( 1 - 2L^2 \laplacian d \right) = 0 & \text{if } \dot{d} > 0,   \vspace{0.5em} \\
          2\,(d-1) W^{e}(\tstrain) + m'(d) \left[(1 - b)p \nabla \cdot \tensor{u} + \grad p \cdot \tensor{u} \right] + \dfrac{3\mathcal{G}_{c}}{8 L } \left( 1 - 2L^2 \laplacian d \right) \leq 0 & \text{if } \dot{d} = 0,
    \end{array}
    \right .
        \quad \text{in}  &  \,\, \Omega \times \mathbb{T}
       , \label{eq:damage-eq}\\
     \dfrac{\pd}{\pd t} \left( \phi \rho_{f} \right) + \nabla \cdot  \left(\rho_{f} \tensor{v} \right) - s = 0                 \quad \text{in}  & \,\, \Omega \times \mathbb{T},
\end{align}
subject to the boundary conditions 
\begin{align}
    \tensor{u} &= \tensor{0} & \quad  \text{on} & \quad \pd_{u} \Omega \times \mathbb{T}, \label{eq:disp-bc}\\
    \left[ \tstress'(\tstrain, d) - m(d)\,(b-1) p \, \tensor{1} - p \tensor{1} \right] \cdot \tensor{n}  &=  \hat{\tensor{t}}(\tensor{x}, t) & \quad \text{on} & \quad \pd_{t} \Omega \times \mathbb{T}, \\ \label{eq:traction-bc}
    \grad d \cdot \tensor{n} &= 0  & \quad \text{on} & \quad \pd \Omega \times \mathbb{T},\\
     p &= \hat{p}(\tensor{x}, t) & \quad \text{on} & \quad \pd_{p}\Omega \times \mathbb{T}, \\
    - \tensor{v} \cdot \tensor{n} &= \hat{q}(\tensor{x}, t) & \quad \text{on} & \quad \pd_{q}\Omega \times \mathbb{T}, \label{eq:flux-bc}
\end{align}
and initial conditions
\begin{align}
    \tensor{u}(\tensor{x},0) &= \bar{\tensor{u}}(\tensor{x}) \quad \text{in} \,\, \Omega, \label{eq:disp-initial-condtions} \\
    d(\tensor{x}, 0)         &= \bar{d}(\tensor{x})  \quad \text{in} \,\, \Omega, \label{eq:damage-initial-condtions}\\ 
    p(\tensor{x}, 0)         &= \bar{p}(\tensor{x}) \quad \text{in} \,\, \Omega. \label{eq:pressure-initial-condtions}
\end{align}
Here, $\tstress'(\tstrain, d) := g(d)\,\mathbb{C}^{e}:\tstrain$ is the degraded effective stress tensor, $\phi$ is the rock porosity, $\rho_{f}$ is the fluid density, $\tensor{v}$ is the fluid velocity, and $s$ is the source/sink term (\eg~wells). 
Additionally, $\hat{\tensor{t}}$, $\hat{p}$, and $\hat{q}$ are the prescribed values of traction, pressure, and fluid flux on the boundary, respectively.
Finally, $\bar{\tensor{u}}$, $\bar{d}$, and $\bar{p}$ denote the initial values of the displacement, damage, and pressure, respectively.   

In the equations presented above, the following constitutive relationships are considered.
\paragraph{Porosity}
The porosity is a weighted average of the porosity at the intact rock and that of the fracture, \ie
\begin{align}
    \phi = (1 - d) \phi_{b} + d \phi_{f},
\end{align}
where $\phi_{b}$ and $\phi_{f}$ are the porosity of the intact rock and that of the fracture, respectively. According to the Biot's poroelasticity theory, $\phi_{b}$ is a function of the volumetric strain ($\strain_\text{vol} := \tstrain:\tensor{1}$) and the fluid pressure $p$, \ie
\begin{align}
    \phi_{b} = \phi_{0} + b (\strain_\text{vol} - \strain_{\text{vol},0}) + \dfrac{b-\phi_{0}}{\kappa_{s}} (p - p_{0}). 
\end{align}
Here, $\phi_{0}$, $p_{0}$, and $\strain_{\text{vol},0}$ are the reference values for porosity, pressure, and volumetric strain. From now on, we will assume the porosity of the fracture to be $\phi_{f} = 1$.
\paragraph{Fluid velocity}
The fluid velocity $\tensor{v}$ is computed based on Darcy's law, 
\begin{align}
    \tensor{v} = \dfrac{k}{\mu_{d}} \left( \grad p - \rho_{f} \tensor{g} \right) , 
\end{align}
where $k$ is the rock permeability and $\mu_d$ is the dynamic viscosity of the fluid. 
\paragraph{Fluid properties} 
The fluid density is computed as 
\begin{align}
    \rho_f = \rho_0 e^{\frac{(p-p_0)}{K_f}},
\end{align}
where $\rho_0$ is the density at the reference pressure $p_0$, and $K_f$ is the fluid bulk modulus. The fluid viscosity is considered constant.
\paragraph{Permeability}
The rock permeability $k$ is computed as 
\begin{align}
    k (d) = k_{0}e^{\alpha_{k} d}. \label{eq:perm}
\end{align}
Here, $k_{0}$ is the permeability of the intact material and $\alpha_{k}$ is an empirical coefficient. This relation is introduced based on an experimental measurement of the permeability in diffusely damaged concrete~\cite{pijaudier2009permeability}, and has been previously applied in phase-field simulations~\cite{heider2017phase,pillai2018diffusive,mollaali2019numerical}. A more accurate approach to model the fluid flow in the fracture is to adopt the Reynolds lubrication equation, which relies on the computation of the crack aperture (opening). Unfortunately, existing methods for calculating the fracture aperture have various drawbacks (see \cite{yoshioka2020crack}). Therefore, we adopt Eq.~\eqref{eq:perm} here for simplicity. 

We note that the governing equations presented in this section are derived solely based on the minimization of the total potential energy and do not include any fracture nucleation criterion related to the material strength. As a consequence, these equations are unsuited to model hydraulic fracture nucleation in the bulk material that has no pre-existing fractures. In the next section, we will present how the formulation is extended to include a strength-based fracture nucleation criterion.

% SECTION 3
% ------------------------------------------------------------------------------
\section{A phase-field model for hydraulic fracture nucleation and propagation}
\label{sec:nucleation}

%----------------------------------------------------------------------------------------
%	NUCLEATION MODEL
%----------------------------------------------------------------------------------------

In this section, we extend the phase-field formulation for hydraulic fractures derived in Section~\ref{sec:formulation} to include a stress-based fracture nucleation criterion that takes into account the material strength surface.
In this way, the phase-field method can accurately model not only propagation of preexisting fractures but also hydraulic fracture nucleation in the intact material.

Following the idea introduced by Kumar~\etal~\cite{kumar2020revisiting}, we incorporate the strength surface of the rock material into the model by adding an external driving force term $\hat{c}_{e}(\tstress', L)$ into the damage evolution equation~\eqref{eq:damage-eq}, which gives
\begin{align}
    2(d-1) W^{e}(\tstrain) + m'(d) \left[(1 - b)p \div \tensor{u} + \grad p \cdot \tensor{u} \right] + \hat{c}_{e}(\tstress', L) + \dfrac{3\mathcal{G}_{c}}{8 L } \left[ 1 - 2L^2 \laplacian d \right] = 0 . \label{eq:damage-eq-c_e}
\end{align}
As thoroughly described by Kumar~\etal~\cite{kumar2020revisiting}, the external driving force needs to be designed to reproduce the specific strength surface of the material of interest. Here, for sake of simplicity, we consider the form proposed in~\cite{kumar2020revisiting} for the Drucker--Prager yield surface~\cite{drucker1952soil}, \ie
\begin{align}
    \hat{c}_{e}(\tstress', L) \equiv \hat{c}_{e}(I_1, J_2, L) = \dfrac{1}{1+ \beta_{3} I^2_1 } \left(\beta_{2} \sqrt{J_{2}} + \beta_{1} I_1 + \beta_{0} \right) ,      \label{eq:c_e}
\end{align}
with 
\begin{align}
    \beta_{0} &= \dfrac{3\mathcal{G}_{c}}{8L} \delta^{L} , \\ 
    \beta_{1} &= - \dfrac{3(\stress_\mathrm{cs} - \stress_\mathrm{ts})(1 + \delta^{L})\mathcal{G}_{c}}{16 \stress_\mathrm{ts} \stress_\mathrm{cs} L} - \dfrac{8\mu + 24\kappa - 27\stress_\mathrm{ts}}{144\mu \kappa}(\stress_\mathrm{cs} - \stress_\mathrm{ts}) - \dfrac{\mu+3\kappa}{18\mu^{2}\kappa^2 \mathcal{G}_{c}} \stress_\mathrm{ts}(\stress^3_\mathrm{cs} - \stress^3_\mathrm{ts})L , \\ 
    \beta_{2} &= - \dfrac{9(\stress_\mathrm{cs} + \stress_\mathrm{ts})(1 + \delta^{L})\mathcal{G}_{c}}{16 \sqrt{3}\stress_\mathrm{ts} \stress_\mathrm{cs} L} + \dfrac{8\mu + 24\kappa - 27\stress_\mathrm{ts}}{48\sqrt{3}\mu \kappa}(\stress_\mathrm{ts} + \stress_\mathrm{cs}) + \dfrac{\mu+3\kappa}{6\sqrt{3}\mu^{2}\kappa^2 \mathcal{G}_{c}} \stress_\mathrm{ts}(\stress^3_\mathrm{ts} + \stress^3_\mathrm{cs})L, \\ 
    \beta_{3} &= \dfrac{L\stress_\mathrm{ts}}{\mu \kappa \mathcal{G}_{c}}.
\end{align}
 Here, $\stress_\mathrm{ts}$ and $\stress_\mathrm{cs}$ are the tensile and compressive strengths of a material under uniaxial loading, and $I_{1}$ and $J_{2}$ are the stress invariants. The parameter $\delta^{L}$ in Eq.~\eqref{eq:c_e} is a calibration constant that varies with the material properties, the regularization length, and the mesh spacing. It can be calibrated by simulating crack propagation from existing cracks and ensuring that the results exhibit Griffith-like behavior~\cite{kumar2020revisiting}. 
 
\subsection{Design of the $m(d)$ function}

\begin{figure}[htbp]
    \centering
    \includegraphics[width=0.5\textwidth]{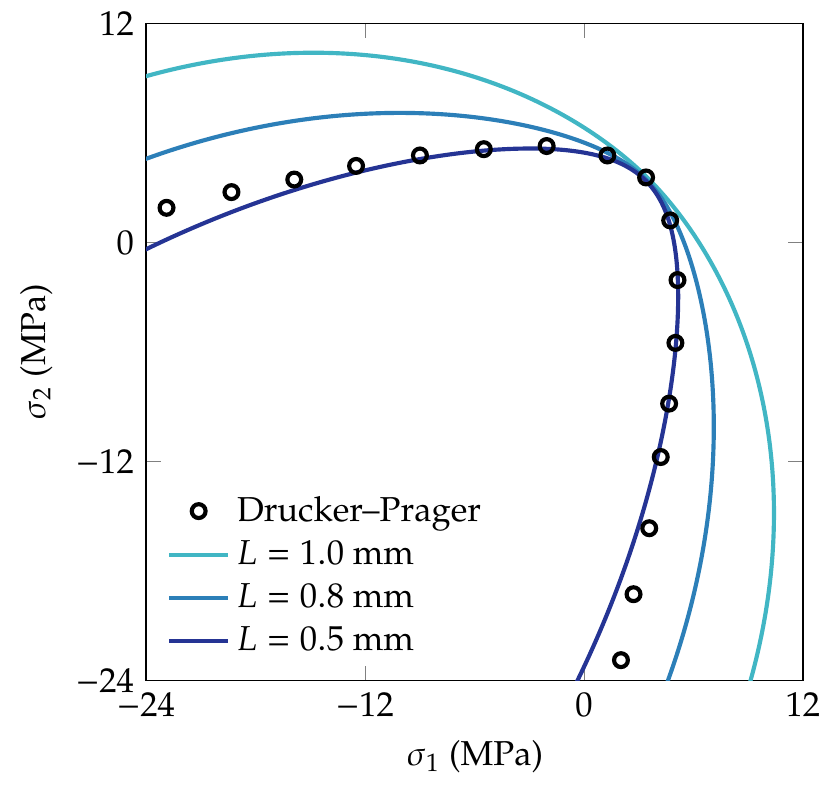}
    \caption{Comparison between the Drucker--Prager model and the strength surfaces predicted by the proposed phase-field method with different regularization lengths.}
    \label{fig:dp-compare}
\end{figure}

For the phase-field model to accurately reproduce the material strength surface, the damage evolution equation for the intact material ($d=0$) should be asymptotically equivalent to the yield surface for $L\rightarrow 0$. 
This is the case for the damage equation below at $d=0$ for modeling non-pressurized fracture in a nonporous material, as mathematically proved in Kumar~\etal~\cite{kumar2020revisiting}, 
\begin{align}
    2W^{e}(\tstrain)- \hat{c}_{e}(\tstress', L) - \dfrac{3\mathcal{G}_{c}}{8L} = 0. \label{eq:damage-eq-at-nucleation}
\end{align}
In our case for modeling hydraulic fractures, however, the damage equation~\eqref{eq:damage-eq-c_e} at $d=0$ as shown below includes pressure terms due to the presence of the fluid,
\begin{align}
    2W^{e}(\tstrain)- m'(0) \left[(1 - b)p \div \tensor{u} + \grad p \cdot \tensor{u} \right] - \hat{c}_{e}(\tstress', L) - \dfrac{3\mathcal{G}_{c}}{8L} = 0.  \label{eq:damage-eq-at-nucleation-pressure}
\end{align}
Thus, to ensure that the damage evolution equation $d=0$ can still asymptotically represent the material strength surface, the damage function $m(d)$ must be chosen such that it satisfies not only Eq.~\eqref{eq:md-requirement} but also $m'(0) = 0$, so that Eq.\eqref{eq:damage-eq-at-nucleation-pressure} becomes identical to Eq.\eqref{eq:damage-eq-at-nucleation}. 
Note that this additional condition ($m'(0) = 0$) is also considered for a similar purpose in Jiang~\etal~\cite{jiang2022phase} to model pressurized cracks in nonporous solids.
To fulfill all these requirements, we choose the following smooth form for $m(d)$ in this work,
\begin{align}
    m(d) = \dfrac{1}{2} \left[1 + \cos(\pi d) \right] .  \label{eq:md}
\end{align}
This choice of $m(d)$ ensures that the strength surface predicted by the phase field model is equivalent to the prescribed one as long as the external driving force is correctly chosen. 
Figure~\ref{fig:dp-compare} shows, for example, a comparison of the strength surface of the Drucker--Prager model and the one predicted by the phase-field model with different regularization lengths, when using the external driving force in Eq.~\eqref{eq:c_e}. 
Here, the phase-field predictions are obtained by analytically solving Eq.\eqref{eq:damage-eq-at-nucleation-pressure} with $m(d)$ defined in Eq.~\eqref{eq:md}.  

\subsection{Enforcement of damage constraints}
For the damage evolution equation~\eqref{eq:damage-eq-c_e} to provide physically meaningful results, it must be subject to two constraints: (1) the boundedness of the damage; (2) irreversibility. In this subsection, we describe how these constraints are enforced in this work.

\paragraph{Boundedness of the damage field}While the damage is guaranteed to be strictly lower than $1.0$~\cite{miehe2016phase}, it is necessary to ensure that the constraint $d\geq0$.  Note that, given an intact material ($d=0$), Eq.~\eqref{eq:damage-eq-at-nucleation-pressure} is only satisfied whenever the stress state reaches the material strength. As a consequence, whenever the stress state of the material is within the strength surface, only a negative damage value satisfies Eq.~\eqref{eq:damage-eq-at-nucleation-pressure}. To avoid such negative damage, we propose to replace the strain energy term in the damage equation by an effective crack driving force term, $\mathcal{D}$, defined as
\begin{align}
    \mathcal{D} = \max \left\{\dfrac{1}{2}\hat{c}^{e}(\tstress', L) + \dfrac{3\mathcal{G}_{c}}{16L}, \, W^{e} (\tstrain) \right\}.
\end{align}
It can easily be seen that this guarantees that an always nonnegative damage solution is obtained even for an intact material.  

\paragraph{Crack irreversibility}Most phase-field models, to avoid crack healing, either employ a history field of the crack driving force~\cite{miehe2010phase,fei2020phase,fei2021double} or an augmented Lagrangian method~\cite{wheeler2014augmented,geelen2019phase} to ensure that the damage is a monotonically increasing function of time, \ie~$\dot{d} \geq 0$ condition in Eq.~\eqref{eq:minimization}. However, as pointed out by Kumar~\etal~\cite{kumar2020revisiting}, the introduction of the external driving force results in a larger diffuse area around the tip of propagating fractures due to the difference between the length-scales of the crack tip and the crack body. Thus, in order to obtain the optimal fracture profile once the damage has fully localized, we only enforce monotonicity once the damage reaches a threshold value, \eg $\,d=0.95$~\cite{kumar2020revisiting}. 
More details of how this constraint is imposed are given in the solution algorithm in Section~\ref{sec:numerical}. 

\subsection{Updated governing equations}
\label{sec:eq-summary}
Given the above modifications, the strong form of the initial boundary value problem is to find the the displacement $\tensor{u}(\tensor{x}, t)$, the damage $d(\tensor{x}, t)$, and the fluid pressure $p(\tensor{x}, t)$ fields that satisfy
\begin{align}
	\nabla \cdot \biggl(\tstress'(\tstrain, d) - m(d) (b-1) p \tensor{1} \biggr)  - m(d) \grad p + \rho \tensor{g} &= \tensor{0} \;\; \text{in} \;\; \Omega \times \mathbb{T} ,  \label{eq:momentum-balance} \\ 
	2(d-1) \mathcal{D} + m'(d) \mathcal{D}_p + \hat{c}_{e}(\tstress', L) + \dfrac{3\mathcal{G}_{c}}{8 L } \left[ 1 - 2L^2 \laplacian d \right] &= 0   \;\; \text{in} \;\; \Omega \times \mathbb{T},  \label{eq:damage-eq-c_e-final} \\ 
	 \dfrac{\pd}{\pd t} \left( \phi \rho_{f} \right) + \diver \left(\rho_{f} \tensor{v} \right) &= s  \;\; \text{in} \;\; \Omega \times \mathbb{T}, \label{eq:flow-eq}
\end{align}
with $\mathcal{D}_p = (1 - b)p \div \tensor{u} + \grad p \cdot \tensor{u}$.
These equations are subject to the boundary conditions in Eqs.~\eqref{eq:disp-bc} - \eqref{eq:flux-bc}, the initial conditions in Eqs~\eqref{eq:disp-initial-condtions} - \eqref{eq:pressure-initial-condtions}, and $\dot{d} \geq 0$ when $d \, \in \, [0.95, 1]$.

% SECTION 4
% ------------------------------------------------------------------------------
\section{Discretization and solution strategy}
\label{sec:numerical}

%----------------------------------------------------------------------------------------
%	NUMERICAL METHOD
%----------------------------------------------------------------------------------------

In this section, we describe the numerical discretization of the governing equations and the solution algorithm.

\subsection{Discretization}
Let us define a mesh $\mathcal{T}$ formed by nonoverlapping cells $K_i$ such that $\Omega \approx \bigcup_i K_i$ and let $\mathcal{F}$ be the set of all faces and $\mathcal{F}_\alpha$ the subset of faces located on $\pd \Omega_\alpha$ with $\alpha = {u, t, p, q}$. Given this mesh, the main unknowns are approximated by their discrete counterparts, \ie~$\tensor{u}^h$, $p^h$ and $d^h$. Additionally, the system of governing equations, Eqs.~\eqref{eq:momentum-balance} - \eqref{eq:flow-eq}, is time-dependent and discretized using discrete time steps $t_i \, \in \, \{t_0, t_1, ..., t_{\max}\}$ and $\Delta t$ indicates the current time interval at $t_{n+1}$, \ie~$\Delta t = t_{n+1} - t_{n}$. In general, the subscripts $(.)_{n+1}$ and $(.)_{n}$ indicate a quantity evaluated at the current time step and the last time step, respectively. From now on, for simplicity, we drop the subscript $(.)_{n+1}$ for the quantity at the current time step. For the spatial discretization, a low order finite element method is employed to discretize the momentum balance~\eqref{eq:momentum-balance} and the damage evolution equation~\eqref{eq:damage-eq-c_e-final}, while the flow equation~\eqref{eq:flow-eq} is discretized by a hybrid mimetic finite difference method~\cite{borio2021}.

Without loss of generality, let us assume homogeneous boundary conditions and define the following three discrete function spaces for the displacement, the damage, and the pressure field, 
\begin{align}
	\mathcal{V}_{u}^{h} & := \left\{ \tensor{\eta}^{h} \,\,  | \,\, \tensor{\eta}^{h} \in [\mathcal{C}^{0} (\overline{\Omega})]^\text{dim}, \,\, \tensor{\eta}^{h} = \tensor{0} \,\, \text{on} \,\, \pd_{u} \Omega,\, \, \tensor{\eta}^{h}_{|K} \in [\mathbb{Q}_{1}(K)]^\text{dim}  \, \, \forall K \in \mathcal{T} \right\},   \\
 	\mathcal{V}_{d}^{h} &:= \left\{ \psi^{h} \,\,  | \,\, \psi^{h} \in \mathcal{C}^{0} (\overline{\Omega}), \, \, \psi^{h}_{|K} \in \mathbb{Q}_{1}(K) \, \, \forall K \in \mathcal{T} \right\},   \\
	\mathcal{V}^{h}_{p} &:= \left\{ \chi^{h} \,\,  | \,\, \chi^{h} \in \mathcal{L}^2(\Omega), \, \, \chi^{h}_{|K} \in \mathbb{P}_{0}(K) \, \, \forall K \in \mathcal{T} \right\}.
\end{align}
Here, $\mathcal{C}^{0} (\overline{\Omega})$ is the space of continuous functions on the closed domain $\overline{\Omega} := \Omega \cup \pd \Omega$, and $\mathbb{Q}_{1}(K)$ is the space of multivariate polynomials on $K$. Additionally, $\mathcal{L}^2(\Omega)$ is the space of square Lebesgue-integrable functions on $\Omega$ and $\mathbb{P}_{0}(K)$ the space of piece-wise constant functions on $K$.  
To discretize the mass balance equation~\eqref{eq:flow-eq}, we also require discretization of the fluxes between neighboring elements. So we define another discrete space $\mathcal{L}^h$ containing the discrete approximation of the face pressure average, ${\pi}^{h} = (\pi_{f})_{f\in \mathcal{F}} \in \mathcal{L}^{h}$, where
\begin{align}
	 \pi_{f} \approx \frac{1}{|f|}\int_{f}p,
\end{align}
 for all $f \in \mathcal{F}$. Note that $\pi_f$ is a face-centered degree of freedom approximating the average pressure on a face. Thus,  we can compute the one-sided face flux, $F_{K,f}$, on the face $f \in \mathcal{F}_K$,  where $\mathcal{F}_K$ is the set of faces in the element $K$, as
\begin{align}
	F_{K,f} = \dfrac{\rho_{f}^\text{upw}}{\mu_d^\text{upw}} \sum_{f' \in \mathcal{F}_{K}} \Upsilon_{ff'} \left[ p_{K} - \pi_{f} - \rho_{f, K} \tensor{g} \cdot (\tensor{x}_{K} - \tensor{x}_{f}) \right].
\end{align}
Here, $\rho_{f}^\text{upw}$ and $\mu_d^\text{upw}$ are the upwinded fluid density and viscosity. Additionally, $\tensor{x}_{K}$ and $\tensor{x}_{f}$ are the locations of the element and face centers, respectively, and $\Upsilon$ is the local transmissibility matrix, evaluated using the quasi two-point flux approximation (TPFA) (see Chapter 6 of~\cite{lie2019introduction} for more details). 

Thus, the discrete weak form of the problem is: find $\{\tensor{u}^{h}, d^{h}, p^{h}, \pi^{h}\}_{n+1} \, \in \, \mathcal{V}_{u}^{h} \times \mathcal{V}_{d}^{h} \times \mathcal{V}_{p}^{h} \times \mathcal{L}^{h}$ such that
\begin{align}
	\mathcal{R}^{h}_{u} &= \int_{\Omega^{h}} \grad^{s} \tensor{\eta}^{h} : \tstress \: \dd V + \int_{\Omega^{h}} m(d) \tensor{\eta}^{h} \cdot \grad p \: \dd V- \int_{\Omega^{h}} \rho \tensor{\eta}^{h}\cdot \tensor{g} \: \dd V + \int_{\pd_{t} \Omega^{h}} p \tensor{\eta}^{h} \cdot \tensor{n} \: \dd A = 0, \label{eq:discrete-momentum-balance}\\ 
	\mathcal{R}^{h}_{d} &= \int_{\Omega^{h}} \psi^{h} \left[2(d - 1) \mathcal{D} + m'(d) \mathcal{D}_{p} + \hat{c}_{e}(\tstress', L)  \right]  \: \dd V + \int_{\Omega^{h}} \dfrac{3\mathcal{G}_{c}}{8L}\left(\psi^{h} + 2L^2 \grad \psi^{h} \cdot \grad d^{h} \right) \: \dd V = 0, \label{eq:discrete-damage-eq}\\
    \mathcal{R}^{h}_{p} &:= \int_{\Omega^{h}} \chi^{h} \dfrac{\phi \rho_{f} - \phi_{n} \rho_{f,n}}{\Delta t} \: \dd V + \sum_{K \in \mathcal{T}} \chi^{h}_{K}\left( \sum_{f \in \mathcal{F}_{K}} |f| F_{K,f} \right) - \int_{\Omega^{h}} \chi^{h} s\: \dd V = 0, \label{eq:discrete-flow-eq}\\
    \mathcal{R}^{h}_{c}&:= -\sum_{K \in \mathcal{T}} \sum_{f \in \mathcal{F}_{K}} |f| F_{K,f} \lambda_f + \sum_{f \in \mathcal{F}_q} |f| \hat{q} \rho \lambda_f = 0, \label{eq:discrete-continuity}
\end{align}
for all $\{\tensor{\eta}^{h}, \psi^{h},  \chi^{h}, \lambda^{h}\} \in \mathcal{V}_{u}^{h} \times \mathcal{V}_{d}^{h} \times \mathcal{V}_{p}^{h} \times \mathcal{L}^h$. Note that the discretized continuity equation~\eqref{eq:discrete-continuity} is introduced to ensure continuity of fluxes across faces. The reader may refer to Borio~\etal~\cite{borio2021} for more details on the MFD discretization.
The choice of the MFD method over a more common finite-volume (FV) scheme, is dictated by the need of evaluating the pressure gradient in both the momentum balance~\eqref{eq:momentum-balance} and the damage evolution equation~\eqref{eq:damage-eq-c_e-final}, which is challenging in the standard FV discretization. Here, we take advantage of the additional face-centered pressures to locally approximate the pressure field within an element by using least squares fitting. 

\subsection{Solution strategy}

Equations~\eqref{eq:discrete-momentum-balance} - \eqref{eq:discrete-continuity} form a coupled system of nonlinear equations. To solve these equations, we employ a sequentially-coupled approach, originally proposed by Miehe~\etal~\cite{miehe2010phase} and widely applied in other phase-field literature~\cite{borden2012phase,geelen2019phase,fei2021double}. The solution algorithm is summarized in Algorithm~\ref{algo:sequential-algo}. Given the displacement, pressure, and damage fields at the previous time step $t_n$, $\{\tensor{u}^{h}_n, p_n^{h}, \pi_n^{h}, d_n^{h}\}$, we enter staggered interations, in which the poromechanics system (Eqs.~\eqref{eq:discrete-momentum-balance}, \eqref{eq:discrete-flow-eq}, and \eqref{eq:discrete-continuity}) and the phase-field equation~\eqref{eq:discrete-damage-eq} are solved sequentially. 
Specifically, at each staggered iteration, Eqs.~\eqref{eq:discrete-momentum-balance}, \eqref{eq:discrete-flow-eq} and \eqref{eq:discrete-continuity} are solved first by freezing the damage so as to find $\{\tensor{u}^{h}, p^{h}, \pi^{h}\}$ using a fully-coupled approach. Subsequently, the pressure- and displacement-dependent terms are updated followed by solving Eq.~\eqref{eq:discrete-damage-eq} to find the damage field $d^{h}$. 
The updated damage field is then used in the poromechanics solve for displacement and pressure fields at a new iteration step until the nonlinear solvers of both poromechanics and phase-field converge in just one step.
Note that the nonlinear equations of poromechanics and phase-field are solved using a Newton-Raphson method. 
Finally, the crack irreversibility constraint is enforced by imposing $d=1$ if damage reaches or exceeds $0.95$. 

\begin{algorithm}[htbp!]
    \setstretch{1}
    \caption{Sequentially coupled phase-field hydraulic fracturing simulation}
      \begin{algorithmic}[1]
        \Require $\tensor{u}_{n}^{h}$, $d_{n}^{h}$, $p_{n}^{h}$ and $(\grad p)_{n}^{h}$.
        \Repeat 
        \State Solve Eqs.~\eqref{eq:discrete-momentum-balance}, \eqref{eq:discrete-flow-eq}, and \eqref{eq:discrete-continuity} for $\tensor{u}^{h}$, $p^{h}$, and $\pi^{h}$ freezing $d^{h}$ (poromechanics solve).
        \State Compute element-wise $\grad p$ using $p^{h}$ and $\pi^{h}$.
        \State Update drivng forces $\mathcal{D}$, $\mathcal{D}_{p}$ and $\hat{c}_{e}$. 
        \State Solve Eq.~\eqref{eq:discrete-damage-eq} freezing $\tensor{u}^{h}$ and $p^{h}$ (phase-field solve). 
        \Until{Both poromechancis solve and phase-field solve converge in one Newton iteration. }
        \State If $d \geq 0.95$, $d=1$ (irreversibility constraint). 
        \Ensure $\tensor{u}^{h}$, $d^{h}$, $p^{h}$, and $\grad p^{h}$.
      \end{algorithmic}
  \label{algo:sequential-algo}
\end{algorithm}

% SECTION 5
% ------------------------------------------------------------------------------
\section{Numerical examples}
\label{sec:simulation}

%----------------------------------------------------------------------------------------
%   SIMULATION RESULTS
%----------------------------------------------------------------------------------------

In this section, we present three numerical examples to demonstrate that the proposed phase-field model can accurately reproduce hydraulic fracture nucleation and propagation and to illustrate its applicability to relevant physical scenarios. 
The first two 2D examples are designed to demonstrate that the proposed method is able to (i) correctly model fracture propagation, given a properly calibrated $\delta^{L}$, and (ii) accurately predict hydraulic fracture nucleation in the bulk due to stress-induced failure.
In the last example, we apply the proposed phase-field method to model hydraulic fracturing in more realistic 3D settings. 

The material parameters used in all numerical examples are presented in Table~\ref{tab:solid-parameters}. 
These are consistent with those presented in the literature~\cite{lamond2006significance,lu2022thermally} for a rock-analog concrete sample.
Note that the gravitational force is neglected in all simulations. 

\begin{table}[htbp]
    \centering
    \begin{tabular}{lccc}
        \toprule
        \textbf{Parameter} & \textbf{Symbol} & \textbf{Value} & \textbf{Unit} \\ 
        \midrule
        Bulk modulus & $\kappa$ & $16.7  $ & GPa \vspace{2pt} \\
        Shear modulus & $\mu$ & $12.5 $ & GPa \vspace{2pt} \\
        Critical fracture energy & $\mathcal{G}_{c}$ & 4 & J/m$^2$ \vspace{2pt} \\
        Tensile strength & $\sigma_\mathrm{ts}$ & 5.5 & MPa \vspace{2pt} \\
        Compressive strength & $\sigma_\mathrm{cs}$ & 40 & MPa \vspace{2pt} \\
        \bottomrule
    \end{tabular}
    \caption{Material properties employed in all numerical examples.}
    \label{tab:solid-parameters}
\end{table}

\subsection{Example 1: pressurized crack propagation in a nonporous solid}
\begin{figure}[htbp]
    \centering
    \includegraphics[width=0.3\textwidth]{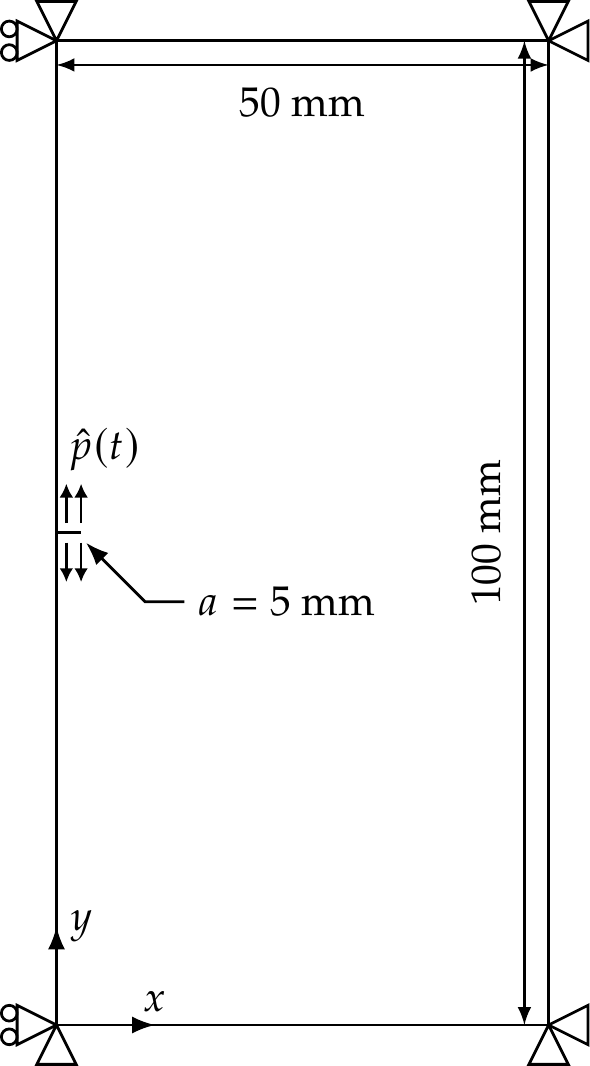}
    \caption{Pressurized crack propagation in a nonporous solid: geometry and boundary conditions.}
    \label{fig:pressure-crack-setup}
\end{figure}

As a first numerical example, we consider the test case proposed by Sneddon and Lowengrub~\cite{sneddon1969crack} and extend it to the propagation of a uniformly pressurized crack in a nonporous elastic solid. This test case has been previously presented in the literature~\cite{bourdin2012variational,wheeler2014augmented,mikelic2015quasi,wilson2016phase,jiang2022phase}. The purpose of this numerical example is to demonstrate how to calibrate the coefficient $\delta^{L}$ to accurately predict fracture propagation consistently with Griffith's theory~\cite{griffith1921vi} and to investigate how the calibrated value varies with the mesh resolution.
The geometry and the boundary conditions are presented in Figure~\ref{fig:pressure-crack-setup}. We consider a $50\, \text{mm} \, \times \, 100 \, \text{mm}$ rectangular nonporous domain (fluid flow is not considered in the bulk). An horizontal crack with length $a= 5 \, \text{mm}$ is present at the center of the left boundary. 
Note that the ratio between the initial crack length and the domain size is sufficiently small to approximate an infinite medium. 
A uniformly distributed pressure, $\hat{p}$, linearly increasing as a function of time (\ie~$\hat{p}(t) = 0.1568 \, [\text{MPa/s}] \cdot t [\text{s}]$) is applied to the fracture until propagation begins. The time step size is set to be $\Delta t = 0.1$ s. 

The critical pressure that triggers fracture propagation can be computed analytically based on the benchmark solution in Sneddon and Lowengrub~\cite{sneddon1969crack} and linear-elastic fracture mechanics (LEFM) theory~\cite{griffith1921vi,irwin1957analysis}, \ie 
\begin{align}
    p_\text{crit}^\text{ref} = \sqrt{\dfrac{E \mathcal{G}_{c}}{(1 - \nu^{2}) \pi a}} = 2.8218 \text{ MPa},
\end{align}
where $E$ is Young's modulus and $\nu$ is Poisson's ratio. We employ this analytical solution as a reference to calibrate $\delta^{L}$. 
The calibrated values of $\delta^{L}$ are listed in Table~\ref{tab:delta-L} for two regularization lengths and various levels of mesh refinement $L/h$, where $h$ denotes the element size. 
Remark that with the properly calibrated $\delta^{L}$ values, the proposed phase-field method retains the capability to correctly model the energy-based fracture propagation process.

\begin{table}[htbp]
    \centering
    \begin{tabular}{l|cccc|cccc}
        \toprule
        Regularization length $L$ (mm) & \multicolumn{4}{c|}{0.5} & \multicolumn{4}{c}{1.0}\\
        \midrule
        Mesh level $L/h$ & 4 & 5 & 8 & 10 & 4 & 5 & 8 & 10 \\
        \midrule
        Calibrated $\delta^{L}$ & 3.31 & 4.15 & 6.62 & 6.85 & 3.28 & 3.65 & 4.16  & 4.48 \\
        \bottomrule
    \end{tabular}
    \caption{Pressurized crack propagation in a nonporous solid: calibrated values of $\delta^{L}$ for each regularization length $L$ and mesh level $L/h$ employed in the simulation.}
    \label{tab:delta-L}
\end{table} 

In Kumar~\etal~\cite{kumar2020revisiting}, it was observed that $\delta^{L}$  grows monotonically as a functions of the phase field regularization length. The same trend is observed in Table~\ref{tab:delta-L} for a pressurized crack. Additionally, given a fixed regularization length, the calibrated value of $\delta^{L}$ increases as the mesh is refined and it converges to a unique value for highly refined meshes.

\begin{figure}[htbp]
    \centering
    \subfloat[$L=0.5$ mm]{\includegraphics[width=0.48\textwidth]{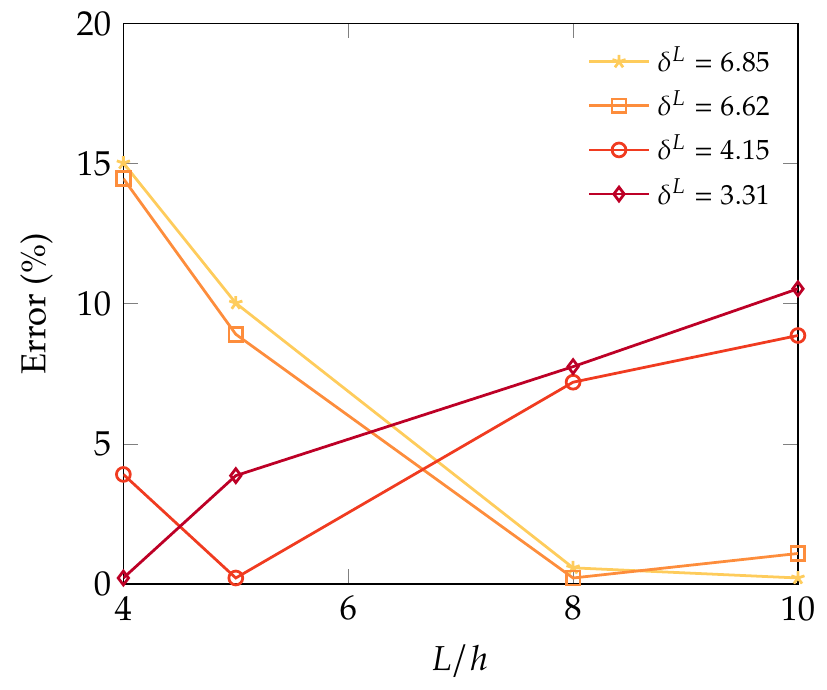} } 
    \subfloat[$L=1.0$ mm]{\includegraphics[width=0.48\textwidth]{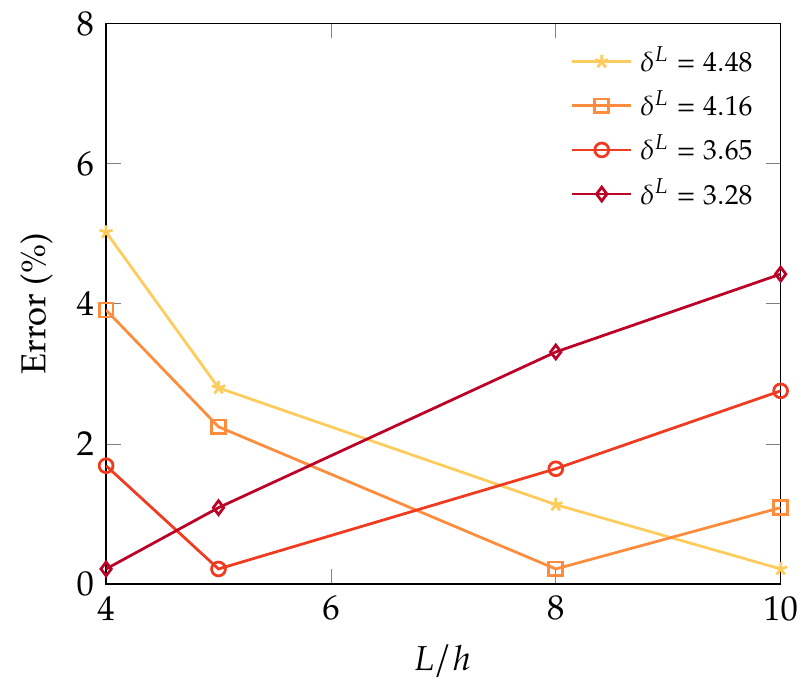}} 
    \caption{Pressurized crack propagation in a nonporous solid: errors of the predicted critical pressure when a fixed $\delta^{L}$ is used at each mesh level for the case with (a) $L = 0.5$ mm, and (b) $L=1.0$ mm.}
    \label{fig:error-plot}
\end{figure}

Since the calibrated value of $\delta^{L}$ is a function of the mesh resolution, undesirable errors may be introduced if a constant $\delta^{L}$ is employed with meshes with nonuniform spacing. Figure~\ref{fig:error-plot} shows the percentage error, computed as $\text{Error} = \frac{p_\text{crit} - p_\text{crit}^\text{ref}}{p_\text{crit}^\text{ref}}$, obtained for each calibrated value of $\delta^L$ as a function of the mesh resolution. Note that larger errors are obtained as the prescribed $\delta^{L}$ deviates from the calibrated value for a given mesh level. 
These observation will support the choice of the appropriate value of $\delta^L$ in the next two numerical examples which employ nonuniform meshes. 

\subsection{Example 2: 2D near-wellbore nucleation and propagation of hydraulic fractures}

\begin{figure}[htbp]
    \centering
    \includegraphics[width=\textwidth]{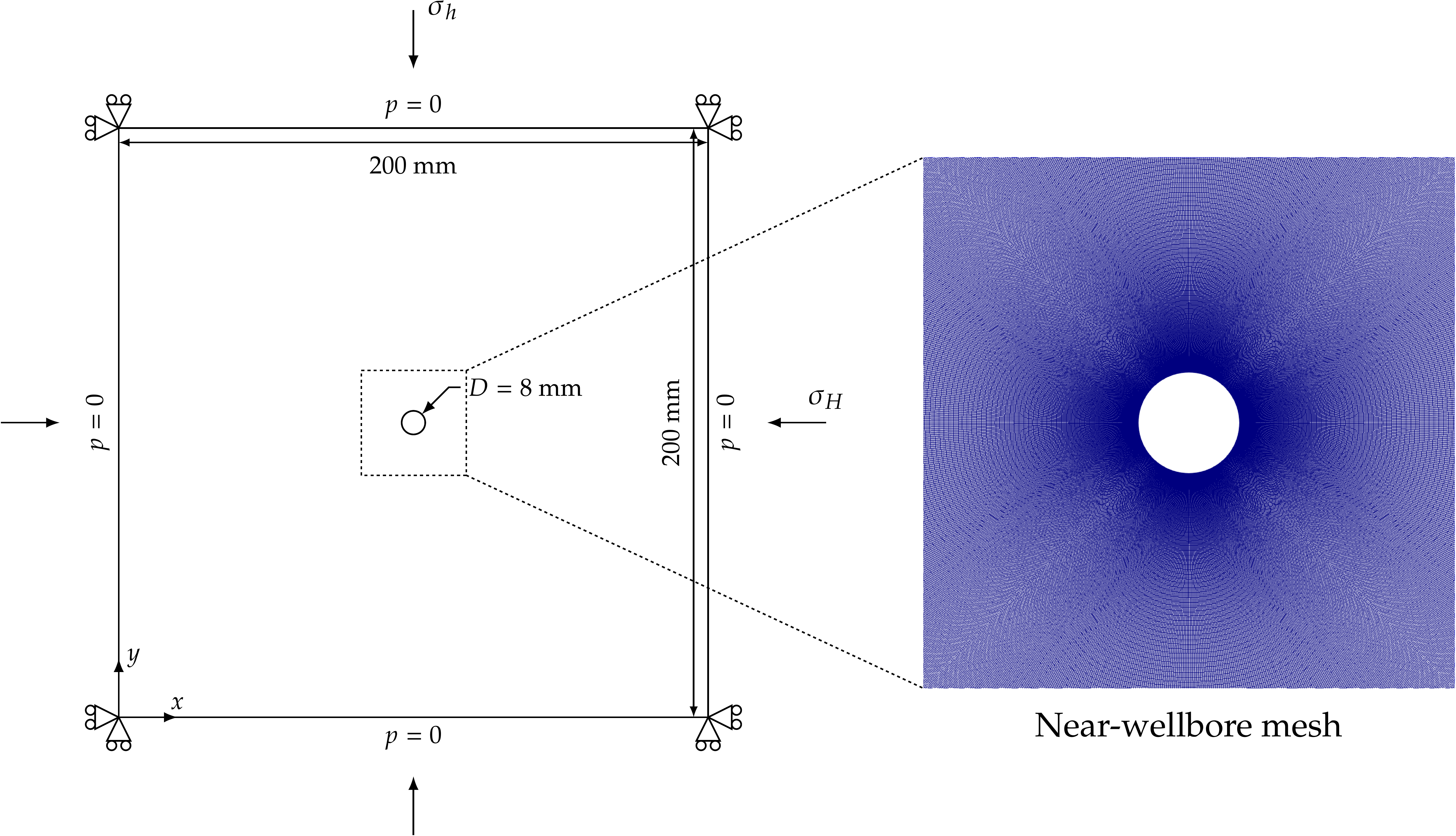}
    \caption{2D near-wellbore nucleation and propagation of hydraulic fractures: geometry, boundary conditions, and near-wellbore mesh.}
    \label{fig:2d-wellbore-setup}
\end{figure}

As a second example, we consider a 2D $200 \, \text{mm} \times 200 \, \text{mm}$ porous domain that contains a circular wellbore at the center with 8 mm in diameter. 
The domain is fully saturated and we simulate the nucleation and propagation of hydraulic fractures in the near-wellbore region due to fluid injection. The dimension and boundary conditions of the problem are illustrated in Figure~\ref{fig:2d-wellbore-setup}. 
Roller boundary conditions are considered at all outer boundaries of the domain along with a Dirichlet pressure condition, $p=0$. An in-situ stress field aligned with the $x$ and $y$ axes is considered. The maximum, $\stress_{H}$, and minimum, $\stress_{h}$, horizontal stresses are aligned with the $x$ and $y$ axes, respectively.  
 
The fluid injection in the wellbore is modeled by a prescribed pressure boundary condition, $\hat{p}_\mathrm{inj}(t)$, on the wellbore surface with $\hat{p}_\mathrm{inj} (t)= 1 \, [\text{MPa/s}] \cdot t [\text{s}]$. 
According to Eq.~\eqref{eq:traction-bc}, we also impose a normal traction with the same magnitude as the injection pressure $\hat{p}_\mathrm{inj}(t)$ on the inner surface of the wellbore to account for the compression applied on the wellbore surface due to fluid injection.  
The poroelastic and fluid properties considered are provided in Table~\ref{tab:flow-parameters}.
The initial values of material deformation and pore pressure are assumed to be zero, \ie~$\strain_{\vol,0} = 0$ and $p_{0} = 0$ MPa. 
Here, we only run one staggered iteration for each loading step to save computational cost. 
To ensure accuracy and numerical stability, the initial time step size is $\Delta t = 0.5$ s and it is then reduced to $\Delta t = 0.02$ s when fracture nucleation starts ($d$ becomes larger than a threshold, \eg~0.95).  
We employ the phase-field regularization length $L$ of 0.5 mm and
discretize the domain by around 2 million hexahedral elements as shown in Figure~\ref{fig:2d-wellbore-setup}. 
Specifically, we ensure $L/h > 4$ in the region where the fracture will propagate to, and the discretization level increases radially to $L/h=10$ around the wellbore. 
For this problem with varying element sizes, the selection of $\delta^{L}$ is challenging as its calibrated value is mesh dependent according to Table~\ref{tab:delta-L}.
Here, for simplicity, we employ a constant $\delta^{L} = 3.31$. Note that, as shown in Figure~\ref{fig:error-plot}, this choice results in relatively small errors for all mesh resolutions considered. Additionally, it results in a wider diffuse area of the crack body~\cite{kumar2020revisiting} which ensures sufficient mesh resolution for the entire crack.

\begin{table}[htbp]
    \centering
    \begin{tabular}{lccc}
        \toprule
        \textbf{Parameter} & \textbf{Symbol} & \textbf{Value} & \textbf{Unit} \\ 
        \midrule
        Bulk modulus of solid grains & $\kappa_{s}$ & 0.8 & - \vspace{2pt} \\
        Fluid density & $\rho_{f}$ & $10^{3}$ & kg/m$^3$ \vspace{2pt} \\ 
        Fluid viscosity & $\mu_{d}$ & $10^{-9}$ & MPa$\cdot$ s \vspace{2pt} \\
        Initial porosity & $\phi_{0}$ & 0.1 & - \vspace{2pt} \\ 
        Matrix permeability & $k_{0}$ & $10^{-9}$ & mm$^{2}$ \\
        Damage coefficient for permeability & $\alpha_{k}$ & 7 & - \\ 
        \bottomrule
    \end{tabular}
    \caption{2D near-wellbore nucleation and propagation of hydraulic fractures: parameters for modeling poroelastic effects and fluid flow.}
    \label{tab:flow-parameters}
\end{table}

\begin{figure}[htbp]
    \centering
    \includegraphics[width=\textwidth]{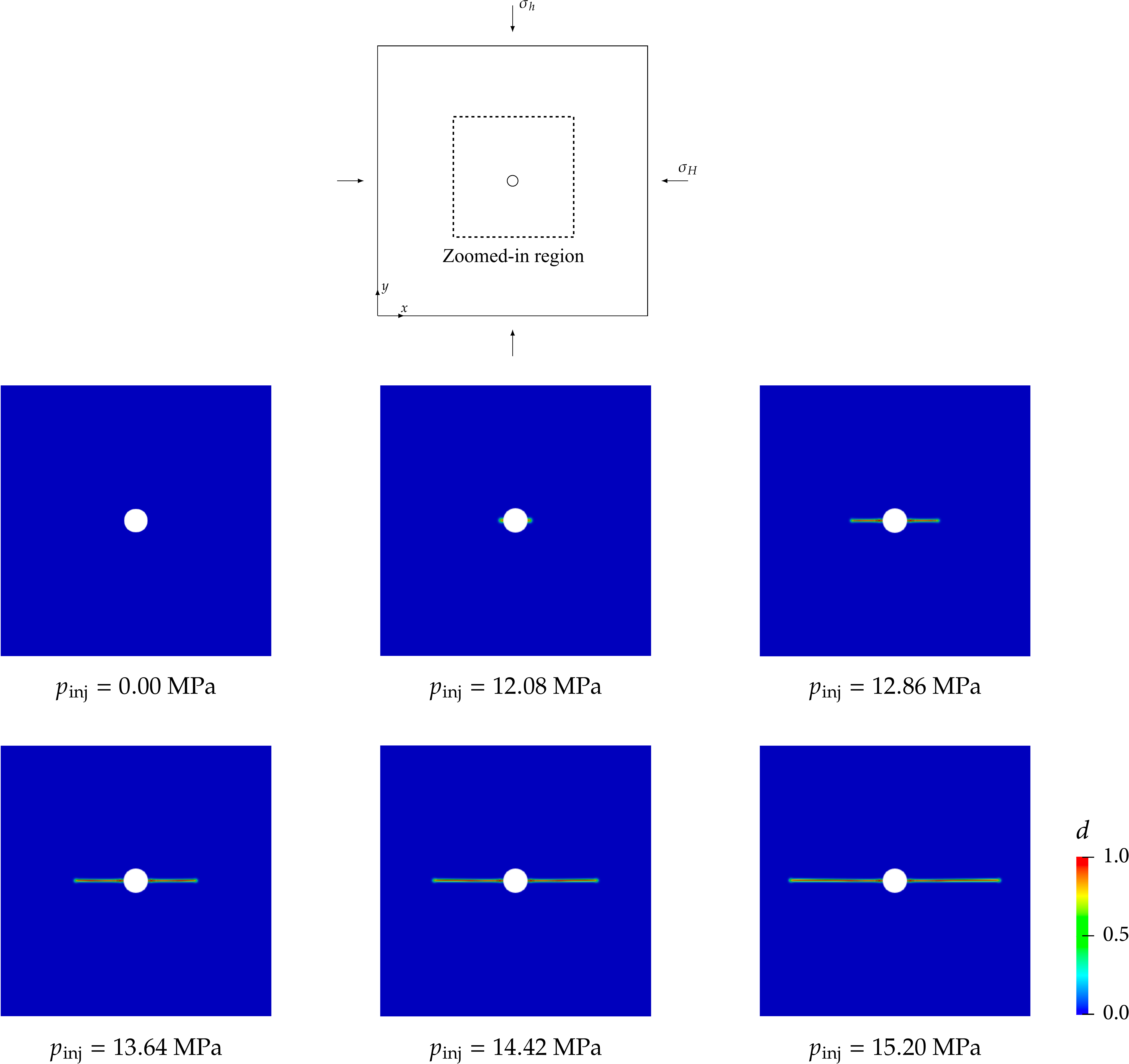}
    \caption{2D near-wellbore nucleation and propagation of hydraulic fractures: phase-field damage evolution.}
    \label{fig:2d-wellbore-damage}
\end{figure}

\begin{figure}[htbp]
    \centering
    \includegraphics[width=\textwidth]{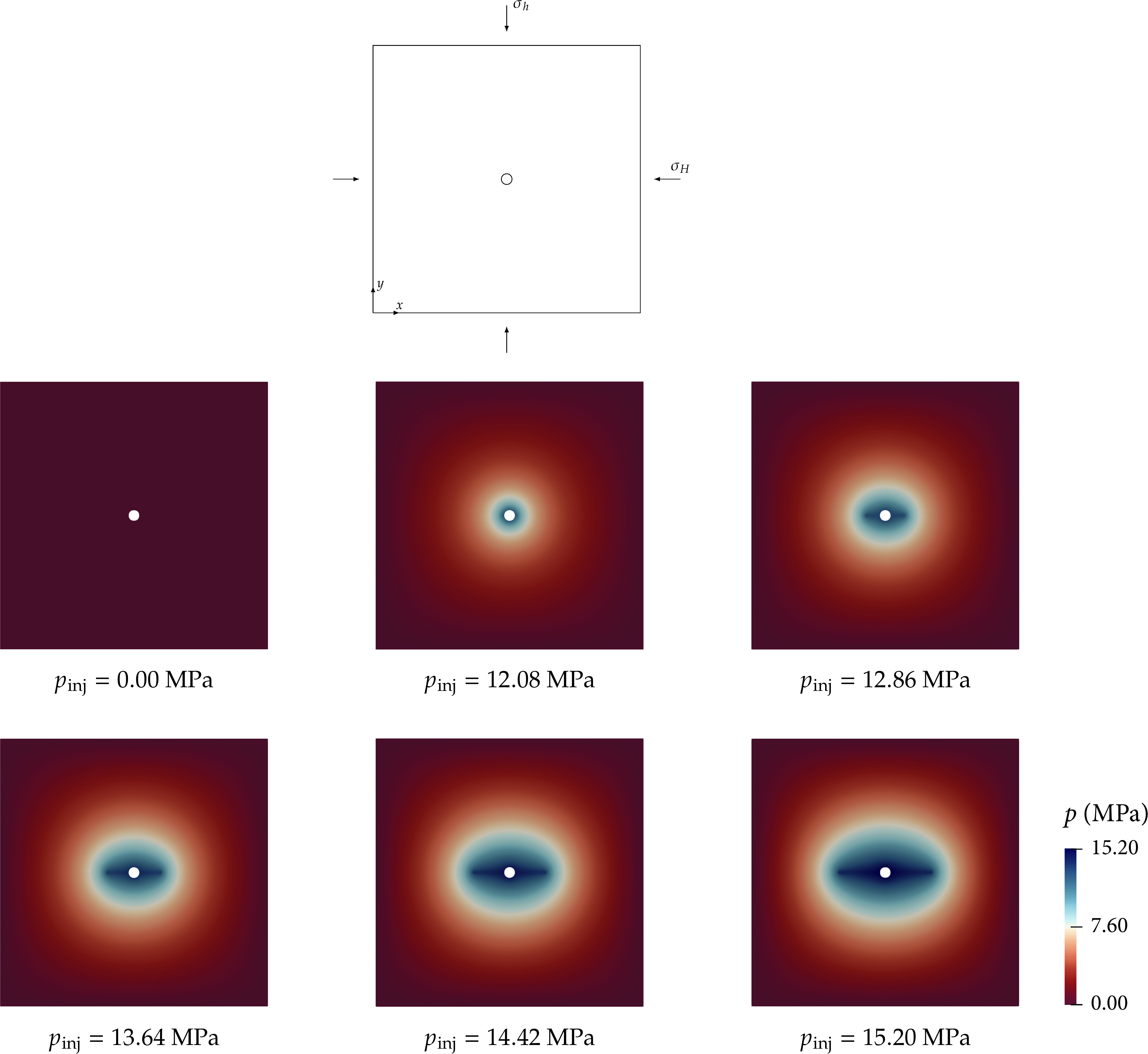}
    \caption{2D near-wellbore nucleation and propagation of hydraulic fractures: pressure evolution.}
    \label{fig:2d-wellbore-pressure}
\end{figure}

As a base case, we consider $\stress_{h} = 8$ MPa and $\stress_{H} = 12$ MPa. Figure~\ref{fig:2d-wellbore-damage} shows the damage field at different simulation stages. 
It is noted that the phase-field method presented in this paper can well capture the nucleation of hydraulic fractures from the smooth wellbore boundary with no preexisting crack/flaws. 
Fracture nucleation occurs due to failure in the bulk material according to the material tensile strength. As the injection pressure increases, the nucleated fractures keep propagating, as expected, in the direction normal to the minimum horizontal stress. Also as shown in Figure~\ref{fig:2d-wellbore-pressure}, the fluid pressure distribution is generally aligned with the fracture propagation direction. This is consistent with the fact that the hydraulic fractures have a higher permeability than the bulk material. As a result, the fluid flow into the fractures is faster than pressure diffusion in the rock matrix. Thus, the fluid pressure builds up inside the fracture, further driving the fracture propagation. 
 
\begin{figure}[htbp]
    \centering
    \includegraphics[width=0.8\textwidth]{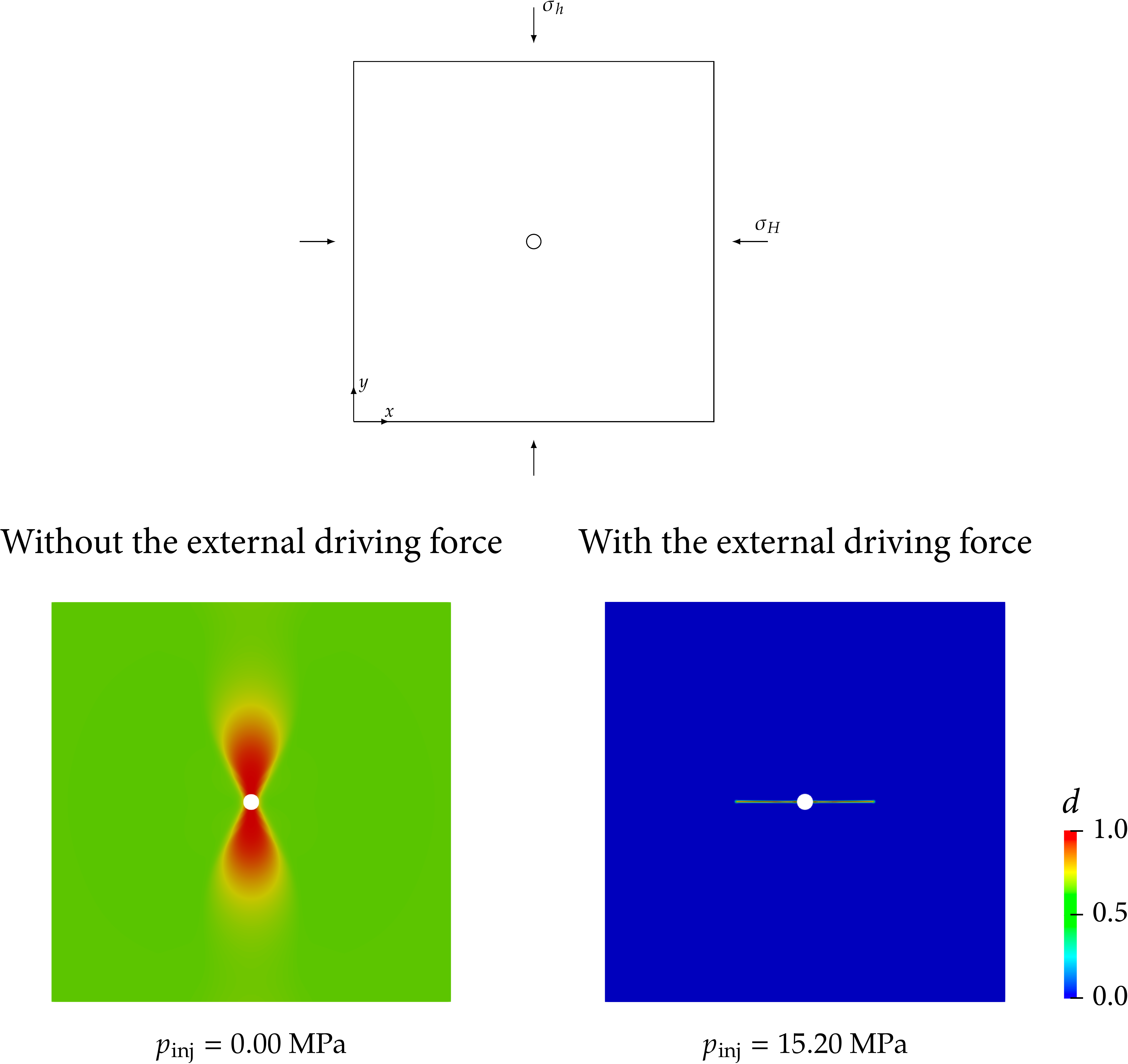}
    \caption{2D near-wellbore nucleation and propagation of hydraulic fractures: comparison of the result produced by the previous phase-field method that has no external driving force (left) and that by the proposed method with the external driving force (right). }
    \label{fig:2d-wellbore-compare}
\end{figure}

We then solve the same test case using a standard phase-field formulation, which does not include the external driving force term. Figure~\ref{fig:2d-wellbore-compare} shows a comparison of the damage field obtained with the two phase-field formulations. Remark that the original phase-field method does not provide the correct direction of damage growth. This can be attributed to its purely energetic formulation, which predicts damage growth whenever the strain energy reaches a certain threshold, while not distinguishing between compressive and tensile strengths. 
Note that a strain energy decomposition (\eg~spectral decomposition~\cite{miehe2010phase}) that only considers the non-compressive component of the strain in the damage equation~\eqref{eq:damage-eq}. This allows to accurately predict the correct fracture propagation pattern. However, even with this strain energy decomposition, the phase-field method requires the calibration of the regularization length $L$ to match the material strength. This calibrated regularization length can be significantly smaller than the problem size, forcing the use of mesh resolution that can easily lead to intractable problem sizes. The phase-field formulation proposed in this paper does not suffer from this limitation due to the $L$-convergence of the strength surface shown in Figure~\ref{fig:dp-compare}. As such, it is an essential extension of the phase-field method for its modeling of near-wellbore hydraulic fracturing.

\begin{figure}[htbp]
    \centering
    \includegraphics[width=0.6\textwidth]{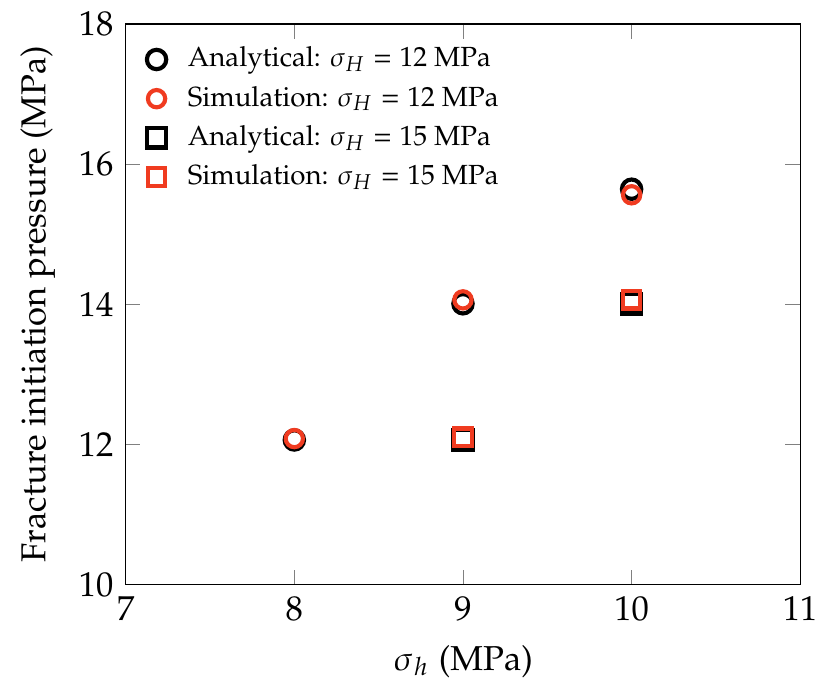}
    \caption{2D near-wellbore nucleation and propagation of hydraulic fractures: comparison between the simulation results and analytical solutions of the fracture initiation pressure under different confining stresses. }
    \label{fig:2d-wellbore-fip}
\end{figure}

Next, we compare the fracture initiation pressure predicted by the phase-field method with that calculated using an analytical model. Here, the fracture initiation pressure in the phase-field simulation is measured as the injection pressure at which the damage reaches the threshold value, 0.95. 
The simulation is performed under five different confining stresses as given in Table~\ref{tab:insitu-stress}.
The reference analytical solution is computed as follows. First, we compute the maximum tangential stress on the wellbore surface based on the analytical model by Haimson and Fairhurst~\cite{haimson1967initiation}, \ie~
\begin{align}
    \stress_{\theta\theta, \max} = \left[2 - \dfrac{b(1 - 2\nu)}{1 - \nu} \right] \hat{p}_\text{inj}(t) - 3 \stress_{h} + \stress_{H} + \dfrac{b(1 - 2\nu)}{1 - \nu} p_{0} . 
\end{align}
Then, given the Drucker--Prager yield function~\cite{drucker1952soil} and the maximum tangential stress $\stress_{\theta\theta, \max}$, we find the radial stress $\hat{\stress}_{rr}$ that equates the yield function to zero, \ie~the material strength is reached.
Since the radial stress $\stress_{rr} = - \hat{p}_\text{inj}$ at all times, we can consider that the fracture initiation pressure is $p_\text{frac} = -\hat{\stress}_{rr}$. 
As shown in Figure~\ref{fig:2d-wellbore-fip}, the phase-field results of the fracture initiation pressure compares very favorably with the analytical solutions for all confining stress combinations, which proves that the proposed method can accurately capture the material strength. Note that the fracture initiation pressure predicted by the phase-field model is not significantly influenced by the choice of $\delta^{L}$. 

\begin{table}[htbp]
    \centering
    \begin{tabular}{l|c|c|c|c|c}
        \toprule
        $\stress_{h}$ (MPa) & 8.0 & 9.0 & 10.0 & 9.0 & 10.0  \\
        \midrule
        $\stress_{H}$ (MPa) & 12.0 & 12.0 & 12.0 & 15.0 & 15.0  \\
        \bottomrule
    \end{tabular}
    \caption{2D near-wellbore nucleation and propagation of hydraulic fractures: confining stresses adopted in the simulation.}
    \label{tab:insitu-stress}
\end{table}

\subsection{Example 3: 3D near-wellbore nucleation and propagation of hydraulic fractures}

\begin{figure}[htbp]
    \centering
    \includegraphics[width=0.6\textwidth]{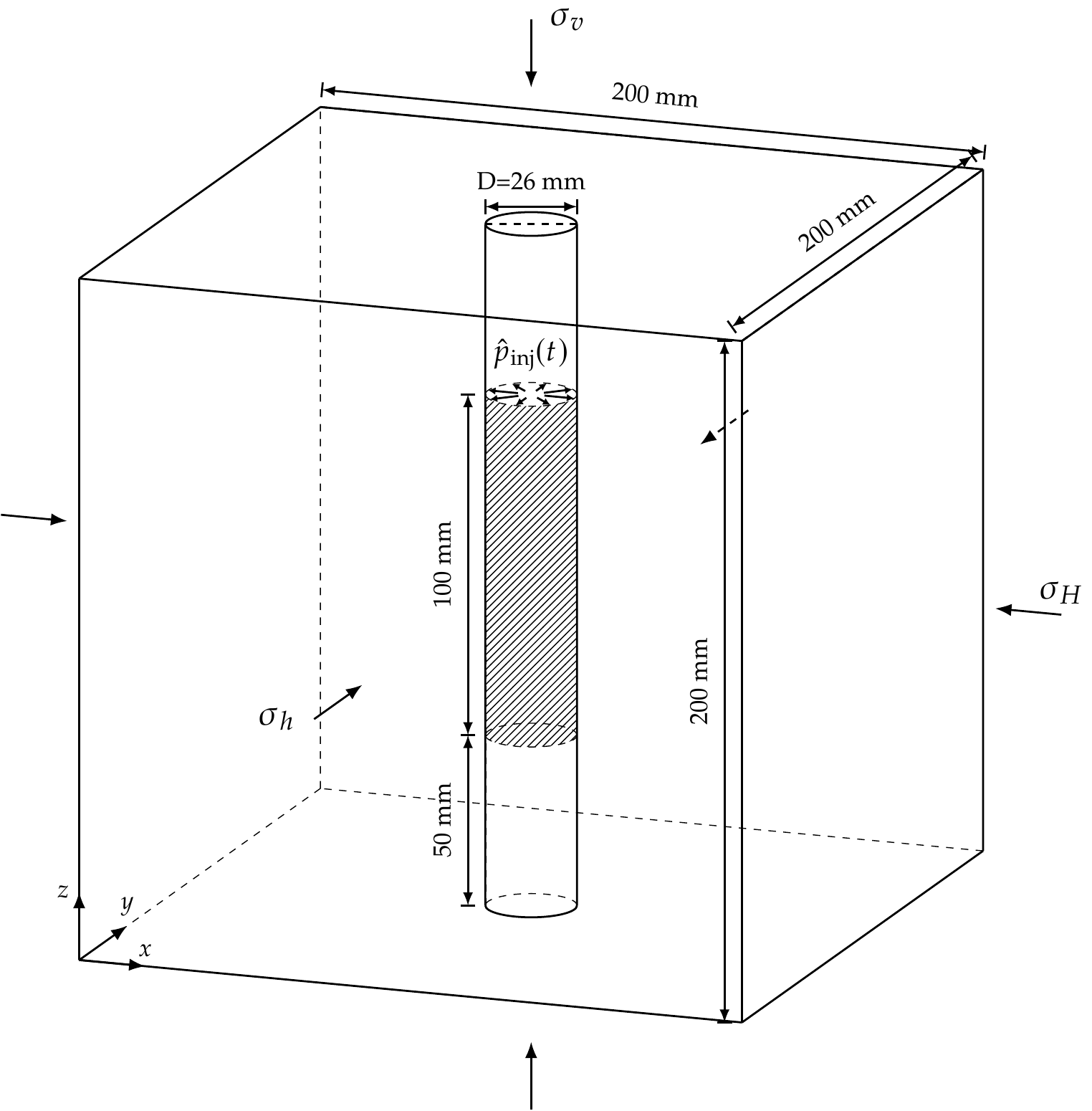}
    \caption{3D near-wellbore nucleation and propagation of hydraulic fractures: geometry and boundary conditions.}
    \label{fig:3d-wellbore-setup}
\end{figure}

In the third example, we simulate a fully 3D near-wellbore hydraulic fracturing problem. 
We consider a $200 \, \text{mm} \, \times \, 200 \, \text{mm} \, \times \,200 \, \text{mm}$ porous domain subject to an in-situ stress field in which principal stresses are aligned with the $x$, $y$, and $z$ directions, as shown in Figure~\ref{fig:3d-wellbore-setup}. 
A vertical (aligned with the $z$-axis) cylindrical wellbore with a diameter of $26$ mm is located in the middle of the domain. Roller boundary conditions are considered on all boundary faces along with a Dirichlet pressure boundary condition, $p=0$.
Additionally, fluid injection is represented by a pressure boundary condition $\hat{p}_\text{inj}(t)$ in the middle section of the wellbore surface (\ie~shaded area in Figure~\ref{fig:3d-wellbore-setup}) with $\hat{p}_\mathrm{inj} (t)= 1 \, [\text{MPa/s}] \cdot t [\text{s}]$. 
The material properties and the time stepping scheme are same as those of the previous 2D wellbore example. 
Single staggered iteration is still adopted for sake of computational time. 
No initial deformation or fluid pressure $p_{0}$ is considered. 
The phase-field regularization length is chosen to be $L = 1$ mm with an element size satisfying $L/h > 4$ near the wellbore, which gives approximately 20 million elements for this problem.
Here, we employ a constant $\delta^{L}$ of 3.28 for the same considerations presented for the previous 2D example. 

We consider two scenarios with different in-situ stress conditions: (i) a uniform minimum horizontal stress $\stress_{h} = 10 \, \text{MPa}$, and (ii) an anisotropic minimum horizontal stress $\stress_{h}$, layered along the vertical direction. This kind of stress distribution has been observed in sedimentary formations due to varying amounts of viscoplastic stress relaxation~\cite{sone2014time,xu2019variation,singh2022predicting,zoback2022lithologically}. Previous studies have shown that the variation of $\stress_{h}$ with depth significantly influences the hydraulic fracture propagation pattern~\cite{xu2019variation,fu2019apparent,zoback2022lithologically}.
Therefore, we employ the proposed phase-field model to numerically investigate the effects of this layered stress condition on the hydraulic fractures pattern.
In both scenarios considered, the vertical stress is $\stress_{v} = 17.5$ MPa, and the maximum horizontal stress (aligned with the $x$-axis) is $\stress_{H} = 15$ MPa. This corresponds to a normal fault stress regime ($\stress_{v} > \stress_{H} > \stress_{h}$). 

\begin{figure}[htbp]
    \centering
    \includegraphics[width=\textwidth]{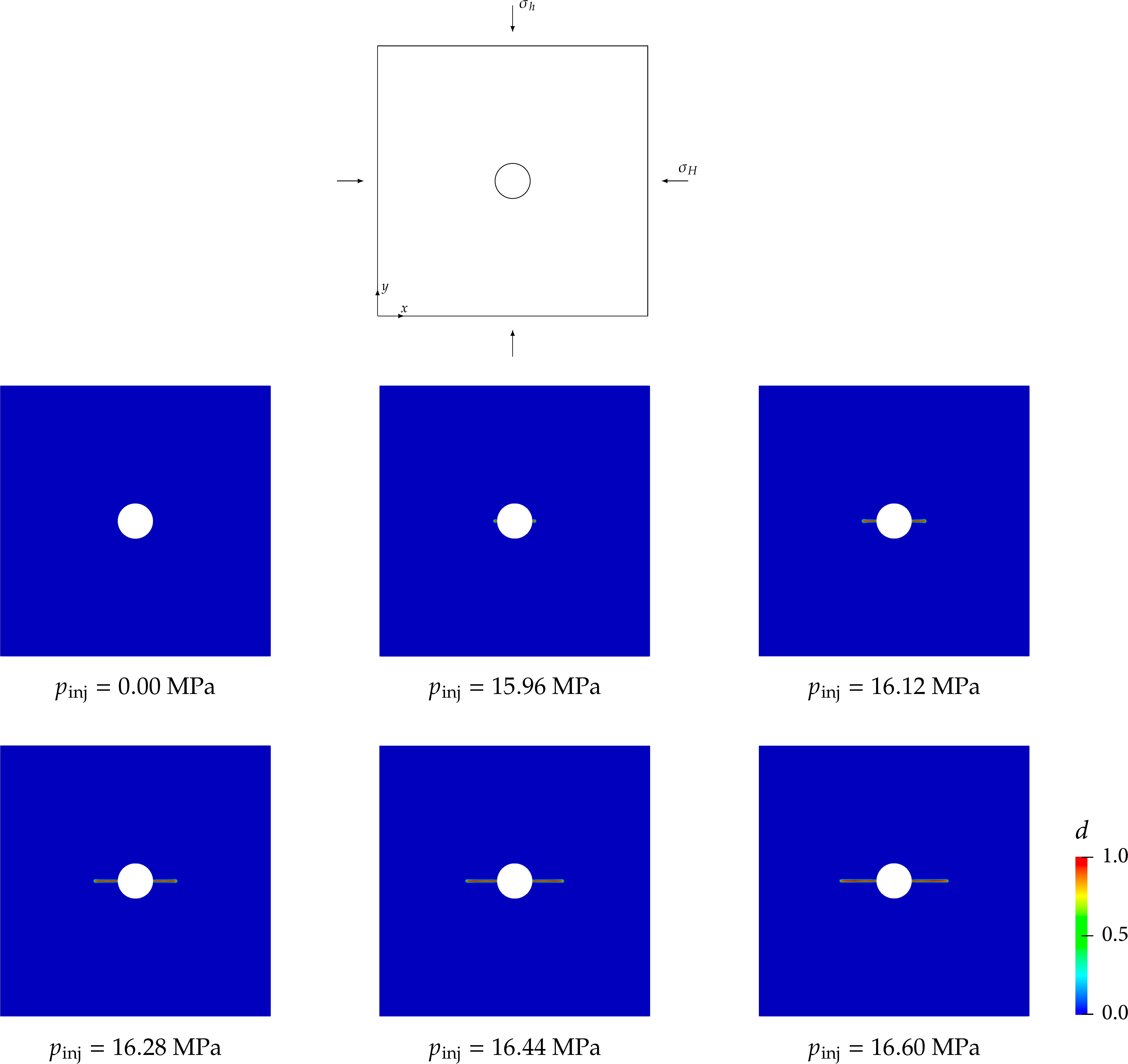}
    \caption{3D near-wellbore nucleation and propagation of hydraulic fractures: 2D view of the phase-field damage evolution at $z = 100$ mm for the case under a uniform $\stress_{h} = 10$ MPa.}
    \label{fig:3d-wellbore-damage-slices}
\end{figure}

\begin{figure}[htbp]
    \centering
    \includegraphics[width=\textwidth]{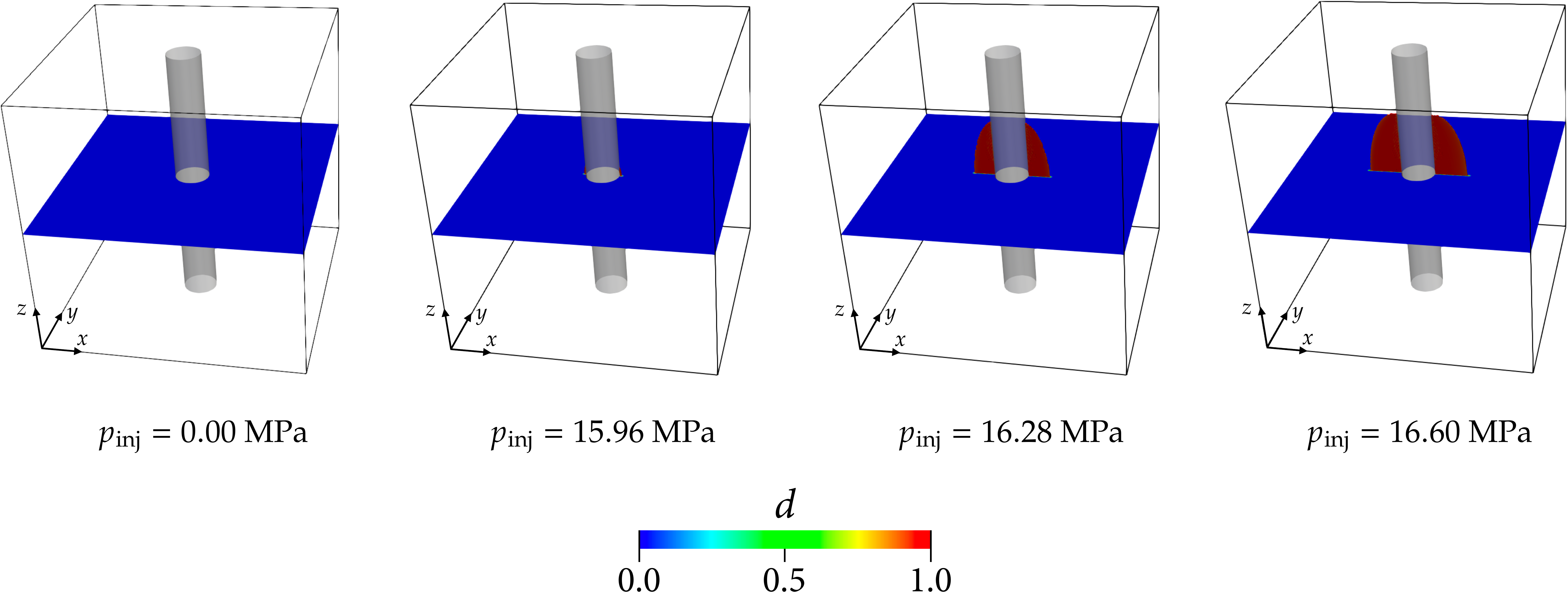}
    \caption{3D near-wellbore nucleation and propagation of hydraulic fractures: 3D view of the phase-field damage evolution for the case under a uniform $\stress_{h} = 10$ MPa.}
    \label{fig:3d-wellbore-damage-volume}
\end{figure}

Figures~\ref{fig:3d-wellbore-damage-slices} and~\ref{fig:3d-wellbore-damage-volume} present a 2D view (at the plane of $z = 100$ mm) and a 3D view, respectively, of the phase-field damage field at different simulation stages for the uniform minimum horizontal stress scenario. Remark that, the nucleation and propagation of bi-wing hydraulic fractures are well captured by the proposed phase-field method. As expected, fractures grow in the $x$-axis which is the direction normal to the minimum horizontal stress. Finally, since injection only occurs in the middle section of the wellbore, hydraulic fractures have an elliptical shape with the maximum propagation distance occuring at the middle plane.  

\begin{figure}[htbp]
    \centering
    \subfloat[]{\includegraphics[width=0.6\textwidth]{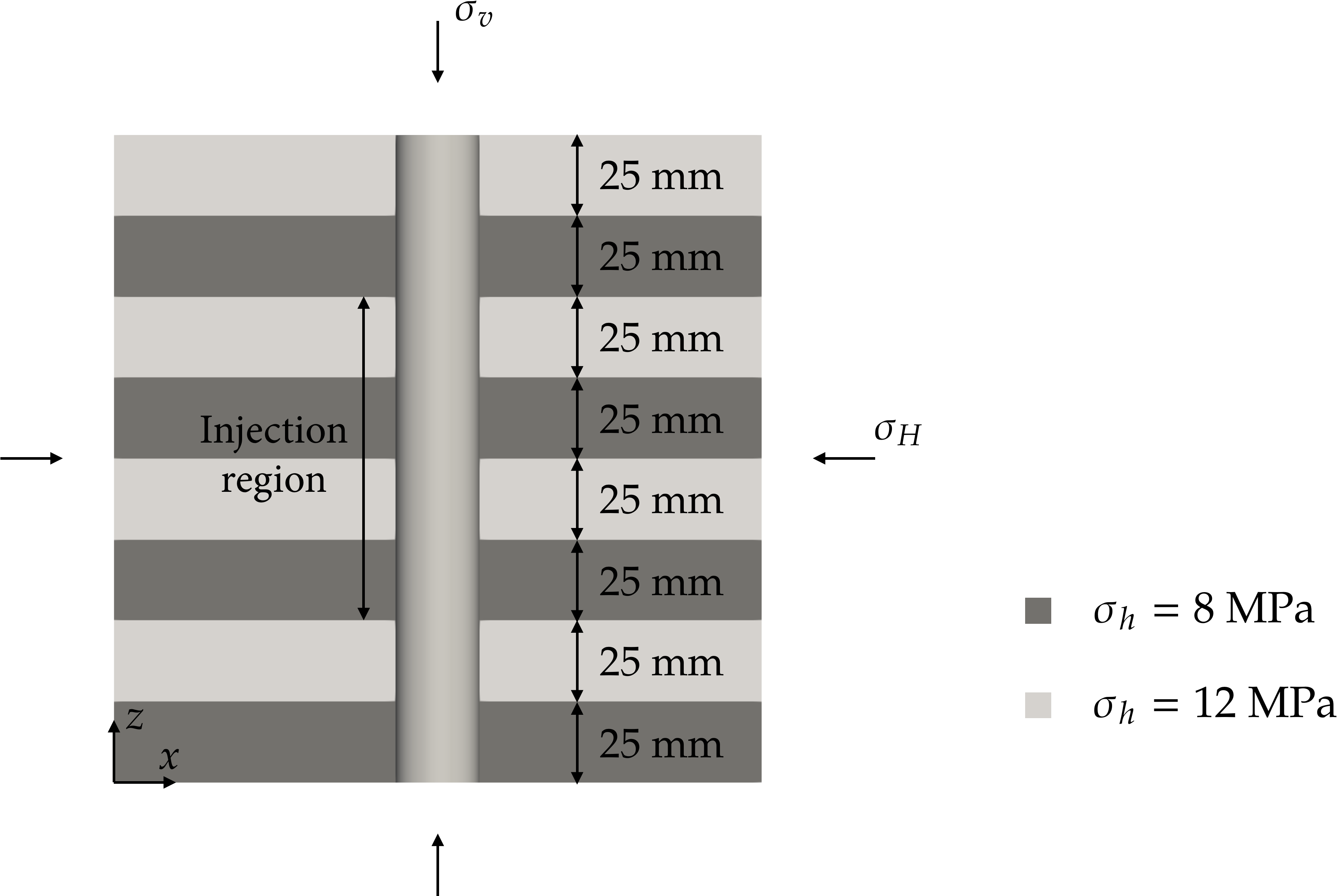} \label{fig:stress-profile}} \hspace{0.5em}
    \subfloat[]{\includegraphics[width=\textwidth]{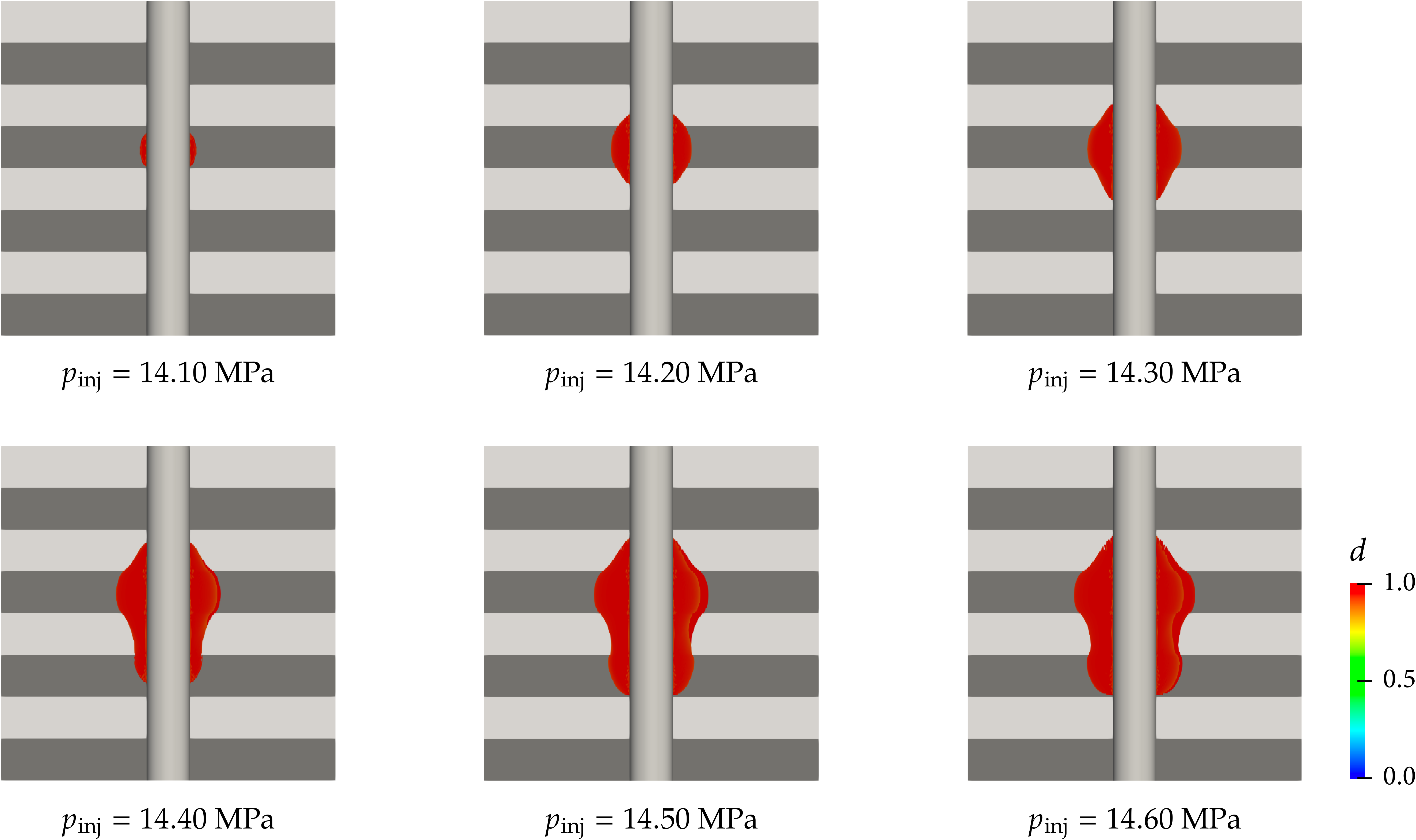} \label{fig:3d-wellboreLayered-damage-slices}} 
    \caption{3D near-wellbore nucleation and propagation of hydraulic fractures: (a) stress profile for the case with alternating high and low $\stress_{h}$ layers; (b) 2D view of the phase-field damage evolution at $y = 100$ mm for the case under alternating $\stress_{h}$.  }
\end{figure}

\begin{figure}[htbp]
    \centering
    \includegraphics[width=\textwidth]{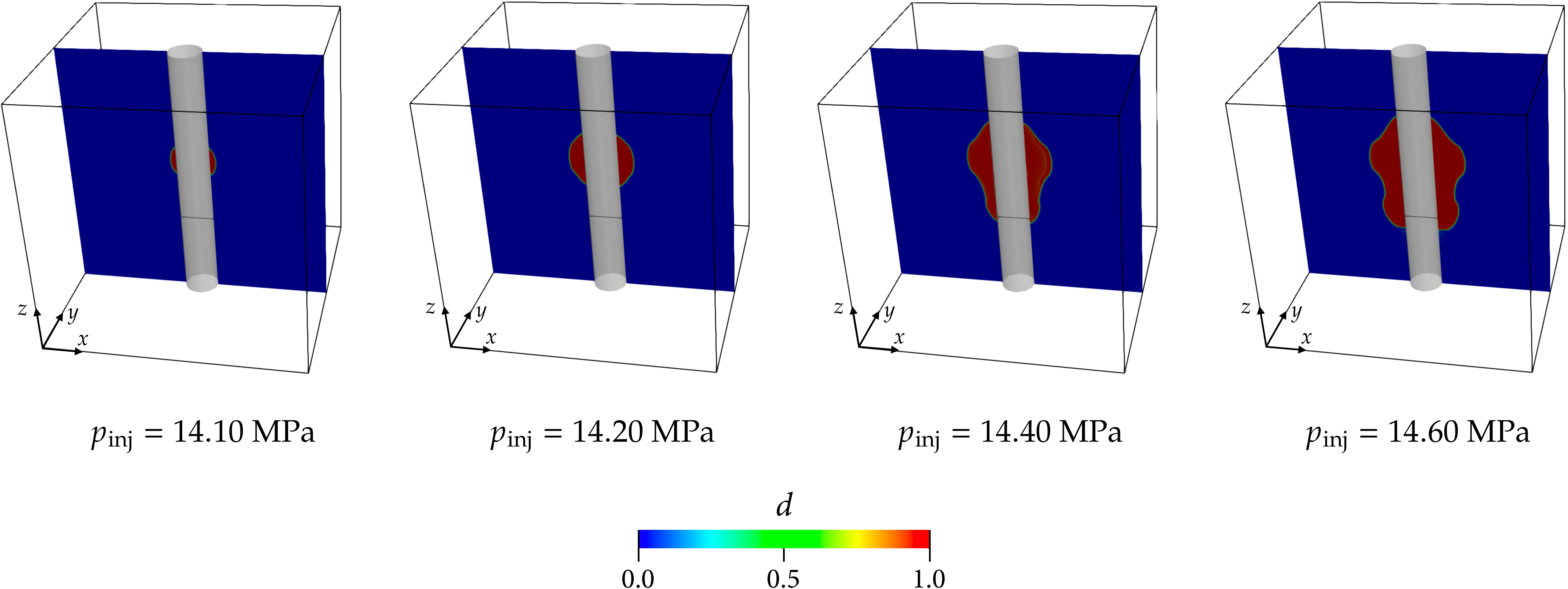}
    \caption{3D near-wellbore nucleation and propagation of hydraulic fractures: 3D view of the phase-field damage evolution for the case under alternating $\stress_{h}$.}
    \label{fig:3d-wellboreLayered-damage-volume}
\end{figure}

Figure~\ref{fig:stress-profile} shows the layered stress distribution considered for the second simulation scenario. This is inspired by the test case presented in Fu~\etal~\cite{fu2019apparent}. The domain is divided into eight $25$ mm high vertical layers, with alternating high-stress ($\stress_{h} = 12$ MPa) and low-stress ($\stress_{h} = 8$ MPa) layers. Figures~\ref{fig:3d-wellboreLayered-damage-slices} and~\ref{fig:3d-wellboreLayered-damage-volume} show a 2D view (at the plane $y =100$ mm) and 3D view of the damage field at different stages of the simulation. Hydraulic fractures first nucleate and propagate in the low-stress layer closer to the injection location. 
Then, they gradually penetrate into the two neighboring layers with a higher stress magnitude.
Subsequently, hydraulic fractures enter the lowest layer within the injection region (the third layer from the bottom to the top). As injection continues, hydraulic fractures keep propagating at a faster rate in the low-stress layers than in the high-stress ones. These hydraulic fracture propagation patterns are consistent with those presented in other numerical investigations~\cite{fu2019apparent}. 

% I WOULD PUT THIS KIND OF CONSIDERATIONS IN THE CONCLUSIONS. Remark that  compared with those results with a uniform $\stress_{h}$ as presented in Fig.~\ref{fig:3d-wellbore-damage-volume}, the simulated hydraulic fractures under alternating $\stress_{h}$ have a more complex shape. This observation thus demonstrates the capability of the phase-field method in handling 3D fracture propagation with arbitrary geometries. 

% SECTION 6
% ------------------------------------------------------------------------------
\section{Conclusions}
\label{sec:conclusion}

%----------------------------------------------------------------------------------------
%	CONCLUSIONS
%----------------------------------------------------------------------------------------

We have proposed a new phase-field approach to hydraulic fracturing in saturated porous media (\eg~rocks) with a speciality in predicting the stress-based characteristics of fracture nucleation. 
Extended from the previous variational formulation, the proposed phase-field method has relied on two key components to incorporate the rock strengths at fracture nucleation in the presence of fluid pressure. 
They are: (i) an external driving force term in the damage evolution equation to account for the material strength, (ii) a special damage function on the crack driving force that is contributed by fluid pressure. 
As verified in numerical examples, the proposed method can accurately capture both energy-based fracture propagation process and stress-based nucleation behavior of hydraulic fractures.
Additionally, we have demonstrated that the phase-field approach can simulate 3D hydraulic fracture nucleation and propagation that involve complex geometries, without requiring sophisticated remeshing algorithms or element enrichment.
We thus believe that the proposed method is an accurate and efficient tool for analyzing and predicting the hydraulic fracturing process in subsurface energy systems. 

Future research will focus on involving more complex features in the proposed phase-field model. 
Examples include the consideration of natural fractures and the coupling with the heat transfer under a non-isothermal condition. 
Such model advancements will undoubtedly foster the numerical investigation of real enhanced geothermal systems.

%----------------------------------------------------------------------------------------
%	ACKNOWLEDGEMENTS
%----------------------------------------------------------------------------------------
\section*{Acknowledgements}
This work relied on the \hyperlink{http://www.geosx.dev/}{GEOS} simulation framework, and the authors wish to thank the GEOS development team for their contributions. 
Funding provided by DOE EERE Geothermal Technologies Office to Utah FORGE and the University of Utah under Project DE-EE0007080 Enhanced Geothermal System Concept Testing and Development at the Milford City, Utah Frontier Observatory for Research in Geothermal Energy (Utah FORGE) site. This work was performed under the auspices of the U.S. Department of Energy by Lawrence Livermore National Laboratory under Contract DE-AC52-07NA27344.  The partial support of  J.\ E.\ Dolbow by NSF grant CMMI-1933367 is also gratefully acknowledged.

%----------------------------------------------------------------------------------------
%	CONTRIBUTIONS
%----------------------------------------------------------------------------------------
\section*{Contribution statement}
\textbf{Fan Fei}: Conceptualization, Methodology, Software, Validation, Formal Analysis, Writing (Original Draft), Visualization. 
\textbf{Andre Costa}: Methodology, Software, Formal Analysis, Writing (Review \& Editing). 
\textbf{John E. Dolbow}: Conceptualization, Methodology, Formal Analysis, Writing (Review \& Editing), Funding Acquisition. 
\textbf{Randolph R. Settgast}: Conceptualization, Software, Funding Acquisition.
\textbf{Matteo Cusini}: Conceptualization, Methodology, Software, Validation, Formal Analysis, Writing (Original Draft), Writing (Review \& Editing), Supervision, Project Administration, Funding Acquisition. 

% REFERENCES
% ------------------------------------------------------------------------------
\bibliography{references}

\end{document}